\documentclass[reprint, prx, superscriptaddress]{revtex4-1}

\usepackage{graphicx}
\usepackage{dcolumn}
\usepackage{bm}
\usepackage{setspace}
\usepackage{booktabs}
\usepackage{amsmath}

\def\half{{1\over 2}}

\usepackage{xcolor}

\def\be{\begin{equation}}
\def\ee{\end{equation}}

\def\1{$|1\rangle$}
\def\2{$|2\rangle$}

\newcommand{\ket}{\rangle}

\newcommand{\re}{\mbox{Re}}

\newcommand{\eexp}{\mbox{e}^}

\begin{document}

\title{Observation of Anomalous Moir\'e Patterns}

\author{Omer Amit}\thanks{equal contribution}
\author{Or Dobkowski}\thanks{equal contribution}
\author{Zhifan Zhou}
\author{Yair Margalit}
\author{Yonathan Japha}
\author{Samuel Moukouri}
\author{Yigal Meir}
\author{Baruch Horovitz}
\author{Ron Folman}\thanks{corresponding author}
	\address{Department of Physics, Ben-Gurion University of the Negev, Be'er Sheva 84105, Israel}



\maketitle

{\bf
Moir\'e patterns are omnipresent. They are important for any overlapping periodic phenomenon, from vibrational and electromagnetic, to condensed matter. Here we show, both theoretically and via experimental simulations by ultracold atoms, that for one-dimensional finite-size periodic systems, moir\'e patterns give rise to anomalous features in both classical and quantum systems. In contrast to the standard moir\'e phenomenon, in which the pattern periodicity is a result of a beat-note between its constituents, we demonstrate moir\'e patterns formed from constituents with the same periodicity. Surprisingly, we observe, in addition, rigidity and singularities. We furthermore uncover universal properties in the frequency domain, which might serve as a novel probe of emitters. These one-dimensional effects could be relevant to a wide range of periodic phenomena.}

Moir\'e patterns are an omnipresent phenomenon\,\cite{review}. They appear when two periodic structures or fields are overlaid. Such patterns may have implications for any overlapping periodic phenomenon. Specifically, in condensed matter, recent years have witnessed the emergence of moir\'e engineering -- the tailoring of electronic, and magnetic properties of van der Waals hetero-structures\,\cite{Zhang2017} or correlated oxides\,\cite{Chen2020}. Such moir\'e materials have also been associated with quantum information and simulation\,\cite{Tran2019,Kennes2021}. Inspired by the subtle role moir\'e patterns may play, we set out to study the formation of these patterns in one-dimensional finite-size periodic systems, and found that such moir\'e patterns give rise to anomalous features in both the classical and quantum domains. In contrast to the standard moir\'e phenomenon, in which the moir\'e-pattern periodicity is a result of a beat-note between the different constituents forming it, we demonstrate moir\'e patterns formed from constituents with the same periodicity. In addition, we observe rigidity and singularities, when varying both the periodicity of the constituents and the relative phase (relative translation). We furthermore uncover universal properties in the frequency domain, which might serve as a novel probe of emitting sources. We simulate such a system with ultracold atoms, precisely controlled by an atom chip\,\cite{keil}, where we make use of a conservation law imposed by the invariance of phase-space distributions under unitary evolution. These one-dimensional effects could be relevant to a wide range of periodic phenomena, including the disciplines of acoustics, optics and solid-state.

The general system we have in mind is a finite-size system in which the envelope of the periodic structure or field is not constant. Such an envelope could be described by a Gaussian, so that the overall sum of the two periodic effects takes the form
\begin{equation}
\begin{split}
V(z)=e^{-(z-z_1)^2/2\sigma^2}\sin^2[\kappa_1(z-z_1)/2+\theta_1]+ \\
e^{-(z-z_2)^2/2\sigma^2}\sin^2[\kappa_2(z-z_2)/2+\theta_2],
\end{split}
\end{equation}
where $\theta_i$ is a phase that determines the position of the maxima (or minima) of the periodic patterns with respect to their Gaussian envelope centers. Unless stated otherwise, we will assume for the constituent wavenumbers $\kappa_1=\kappa_2=\kappa$. We are especially interested in the case $\theta_1=\theta_2$, which may appear in classical waves (e.g. phase-coherent source, or coherent splitting), and is also a fundamental feature of quantum sources, such as that used in our cold-atom simulation [see Supplementary Information (SI)]. In this case, the phase difference between the constituent patterns is proportional to the separation between the two Gaussian peaks $\Delta z=z_2-z_1$, and amounts to $\Delta\phi=\kappa\Delta z$. The finite-size character of the system is best defined by the number of periods, a dimensionless constant proportional to $\kappa\sigma$.

\begin{figure*}
\centering
\includegraphics*[angle=0, width=6in]{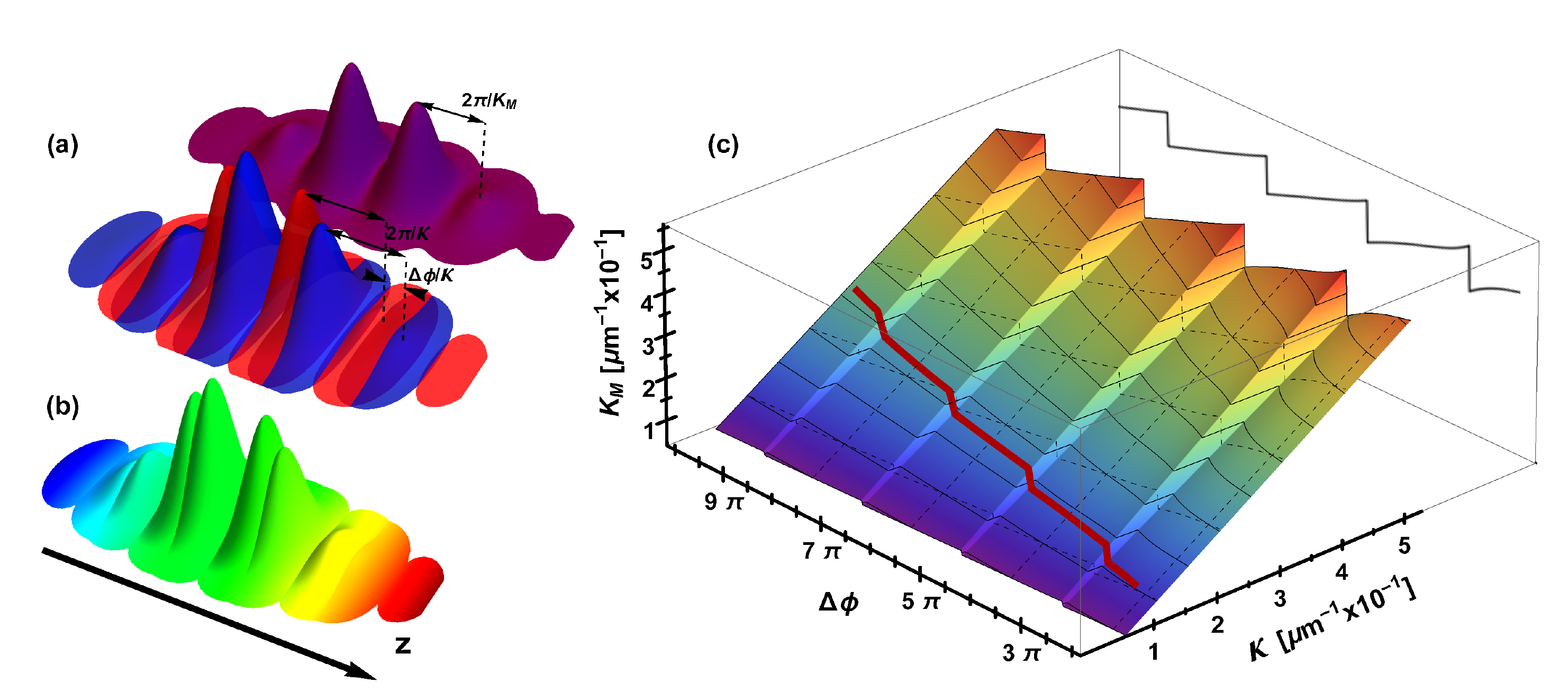}
\centering
\caption[angle=0, width=6in]{Anomalous moir\'e patterns: (a) Two finite-sized one-dimensional periodic structures or fields (blue and red), with wavenumber $\kappa$ and phase difference $\Delta\phi$, join to create a moir\'e pattern (purple) with wavenumber $K_M$. (b) The phase of the moir\'e pattern (color code) projected on the outline of the two constituent periods. The gradient is not constant and has a minimum value at the center, giving rise to a narrow FT peak ($K_M$) which is different from $\kappa$ (see text). At one edge the phase of the blue pattern dominates, as the amplitude of the blue pattern dominates, see (a), and at the other edge the phase of the red pattern dominates. (c) The moir\'e pattern wavenumber $K_M$ (z-axis) vs. $\kappa$ (x-axis) and $\Delta\phi$ (y-axis), as calculated from Eq.\,2. The horizontal dashed grid presents the lines for which $\kappa$ is constant with values of $1-4$. The continuous black line represents the equi-$K_M$ lines ($1.0-5.0$ in $0.5$ increments). For a constant $\kappa$, changing $\Delta\phi$ changes $K_M$. In the background, a projection on the $\kappa=0.45$ plane, assisting to visualize the topography. The two parameters $\kappa$ and $\Delta\phi$ may be made to interplay to ensure the rigidity of $K_M$.
The second (vertical) dashed grid presents the lines for which $\Delta\phi$ equals integer multiples of $2\pi$, identifying the center of the rigidity plateaus presented in Fig.\,2 and Fig.\,4. The red curve represents the actual experimental trajectory giving rise to the data in Figs.\,2 and 4. The fact that this line is always parallel to the equi-$K_M$ lines [expect for the singularity (jump) points], is a manifestation of the moir\'e pattern rigidity. In this model $\kappa\sigma$ is a constant. Specifically, we take the number of periods in the system (edge-to-edge size 4$\sigma$) to have the independently measured value of $N_p=(2/\pi)\kappa\sigma=5.6$. Such joint Fourier measurements on a pair of pulses, may reveal the features of the source, or may simulate solid-state systems (see text). The qualitative features of a corrugated distribution of $K_M$ persist even when the system does not follow the red trajectory (or parallel lines), and even when $\kappa_1$ does not have to exactly equal $\kappa_2$ (Eq.\,1) as in our experiment.
\label{fig:one}}
\end{figure*}

In the first anomalous effect, and in contrast to the known moir\'e phenomenon, the moir\'e periodicity wavenumber, $K_M=2\pi/\lambda_M$, does not result from a beat-note related to $\kappa$, and is instead determined by a translation operation. This is due to a phase gradient which develops along the system. Because of the varying relative magnitude of the two underlying patterns, at one edge of the overall system the phase of the first periodic pattern dominates, while at the other edge, the phase of the second periodic pattern dominates (see the following, as well as SI). This gives rise to a phase gradient which changes $K_M$ as a function of $\Delta\phi$, for any constant $\kappa$ (Fig.\,1).

Adding complexity by varying also $\kappa$, we observe another surprising feature, which is rigidity (i.e., robustness to a change of a central parameter). The observed rigidity is a result of an interplay between the relative phase of the constituents and their period. The effects of these two parameters may in coherent sources naturally cancel each other, thus giving rise to rigidity (quantization) of the moir\'e pattern (SI). Specifically, as shown in Fig.\,1(c) (red trajectory), in our simulated coherent source, when $\kappa$ is changed, the source independently adjusts $\Delta\phi$ to keep $K_M$ constant within certain ranges. Between these ranges, sharp transitions (singularities, jumps) appear in $K_M$. As we show in the SI, at high $\Delta\phi$, where the two envelopes are completely separated, a universal behavior emerges, whereby all the jumps are of the same height.

The phenomena we present are general in the sense that the model may be adapted to different physical scenarios. For example, while Eq.\,1 is always positive, it may just as well describe oscillations between negative and positive values. In addition, the Gaussian envelope used in Eq.\,1 is by no means uniquely suited for this model, and any envelope with a maximum would give similar results. Furthermore, the phenomenon of rigidity and singularities is not crucially dependent on $\kappa$ and $\sigma$ being equal for both constituents (see the following).

Our model may be used to describe the interplay between any two modulated pulses, with a similar modulation frequency and number of periods. These could be, for example, sound or electromagnetic waves, as well as more exotic phenomena. The anomalous features described above may be observed for any phase-coherent source emitting pairs or trains of pulses, as long as there exists a detector with a bandwidth wide enough to follow the oscillations within the pulses.
The model may also be considered as dependent on time rather than space, and could be adapted to describe systems such as Morlet wavelets \cite{Ashmead2012}.

When dependent on space, our model may be used to describe spatially static systems such as those in condensed-matter. Here, a specific application of our model may be found in the context of van der Waals structures, where our model may be seen as a one-dimensional analogue of graphene bi-layer twisting
\cite{Bistritzer2011,Ponomarenko2013,Hunt2013,Dean2013,Shi2014,Gorbachev2014,Song2015,Zhang2017,Spanton2018,Cao2018,Cao2018b,Jin2019,Alexeev2019,Tran2019,Chen2020,Brotons2020,Deilmann2020,Stepanov2020,Nuckolls2020,Zondiner2020,Andrei2020}, where in our experimental simulation
both the periodicity of each of the constituents (equivalent to the lattice constant in the van der Waals structures) and the relative phase (analogous to the twist angle) may be tuned continuously.
A solid-state system could, for example, be a three-layer system, whereby the external layers are periodic structures, such as electrode arrays, which interact with the middle layer (e.g., a 2D gas of electrons). We consider a finite interaction region with varying interaction strength. This could for example be realized with a Gaussian voltage profile on the electrode array. This interaction Hamiltonian thus results in a position-dependent potential term for the middle layer, which is proportional to $V(z)$ (Eq.\,1). $\Delta\phi$ may be manipulated by, for example, having a very short periodicity for the electrodes and imposing the interaction wavenumber $\kappa$ through the voltage profile, and then simply applying a translation of the voltage profile. Self-assembled structures such as graphene layers might also be made to follow the above model. The peaked envelope could probably be manufactured by a varying distance between the layers, or a controllable screening.

We conduct the simulation by utilizing two parallel atom interferometers, each creating its own interference pattern with the form $\exp(-z^2/2\sigma^2)\sin^2(\kappa z/2)$. We then overlap the two interference patterns. We show that as a function of the relative translation $\Delta\phi$, $K_M$ is different from $\kappa$. In addition, we continuously vary $\kappa$ as well as $\Delta\phi$, and we observe frozen moir\'e patterns, for which $K_M$ is not dependent on changes in the parameter $T_2$ (Fig.\,2), the central parameter in our experiment (Fig.\,3), simultaneously determining both $\kappa$ and $\Delta\phi$ (we can also scan $\Delta\phi$ independently by varying $T_3$, and $N_p$ by varying $T_1$). Finally, we note that in our simulation, the number of periods which defines the finite-size character of the system, is naturally fixed (while $\kappa$ and $\sigma$ vary) by a unique conservation law described in detail in the following.

The experimental procedure is depicted in Fig.\,3. Our experiment begins by releasing a Bose-Einstein condensate (BEC) of about $10^4$ $^{87}$Rb atoms from a magnetic trap below an atom chip. We initially prepare the BEC in the state $\vert F, m_F\rangle =\vert 2,2\rangle$, and then create a superposition of the two spin states $|F,m_F\rangle=|2,2\rangle \equiv |2\rangle$ and $|2,1\rangle \equiv |1\rangle$ by applying a $\pi/2$ radio-frequency (RF) pulse. These two states constitute an effective two-level system, as all other states in the $F=2$ manifold are pushed out of resonance by the non-linear Zeeman shift generated using an external bias field (see SI for more details). A Stern-Gerlach interferometer (SGI) is then implemented by using a series of two magnetic gradient pulses (gradients along the axis of gravity, $z$), which are generated by running currents on the atom chip (more details on the setup can be found in Ref.\,\cite{Margalit2019}). The first gradient pulse, of fixed duration $T_1=4\,\mu$s, splits the superposition into two momentum components, which then freely propagate during a delay time $T_{d}$. The wavepackets are then decelerated relative to each other (pulse duration $T_2$). As shown in Fig.\,3, the atoms are exposed to a second $\pi/2$ RF pulse during the delay time $T_{d}$. While the first gradient pulse gives rise to spin-dependent forces, the second gradient decelerates the relative motion of the spin $|1\rangle$ wavepackets due to the non-linear nature of the gradient and the fact that the two wavepackets are at different positions. Due to the differential force the two spin-$|2\rangle$ wavepackets are driven away from the experimental region and are ignored in this experiment. Next, after the deceleration pulse, we apply another $\pi/2$ RF pulse, creating two spin-$|2\rangle$ and two spin-$|1\rangle$ wavepackets, contained within two Gaussian envelopes, all together giving rise to one spin-$|2\rangle$ interference pattern and one spin-$|1\rangle$ interference pattern. As shown in SI, our initial assumption $\theta_1=\theta_2$ (Eq.\,1) is fulfilled. The two patterns now overlap to create a moir\'e pattern, as our imaging is invariant to the spin state.

\begin{figure}
\centerline{
\includegraphics*[angle=0, width=\columnwidth]{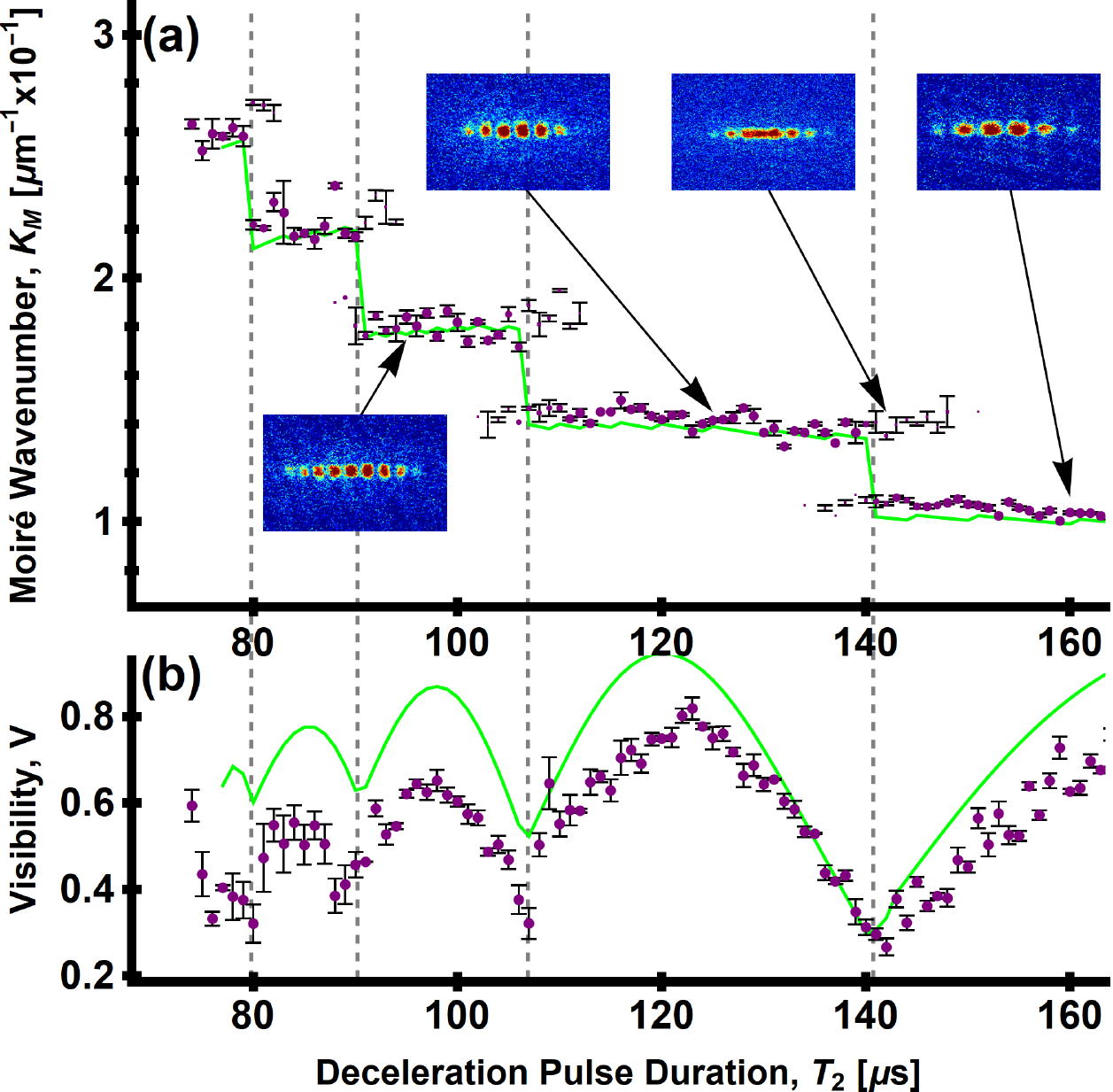}}
\caption{Frozen moir\'e interference pattern periodicity, rigidity and jumps: (a) The measured moir\'e wavenumber $K_M$ [purple, as in Fig.\,1(a)] vs. the deceleration pulse duration, $T_2$. The absolute value of the Fourier Transform (FT) of the moir\'e patterns (CCD images shown in the insets) is calculated from the data, and the value of the maxima, $K_M$, is presented. For values of $T_2$ where a secondary peak is detected with the relative intensity of at least 20\%, two points are plotted with the dot size representing the relative intensity of each peak. The error bars are calculated as the SEM over several iterations. As can be seen, although we keep $\sigma\kappa$ constant, more fringes appear in the CCD insets as $T_2$ becomes smaller. This is due to the growing $\Delta\phi$ which is tantamount to a decreasing ovelap between the two constituent patterns. (b) Visibility of the interference pattern, V, vs. $T_2$. The minima in the visibility, emphasized by the vertical dashed gray lines, correspond to the periodicity jump locations in (a). In both (a) and (b) the green line represents the results of a complete numerical simulation based on the exact experimental conditions. For low values of $T_2$, the simulation overshoots the observed visibility, as under these conditions $2\pi/\kappa$ is smaller, and as the clouds are moving (free-fall), the limited optical resolution gives rise to smearing and consequently a smaller visibility.
\label{fig:data}}
\end{figure}

\begin{figure}
\centerline{
\includegraphics*[angle=0, width=\columnwidth]{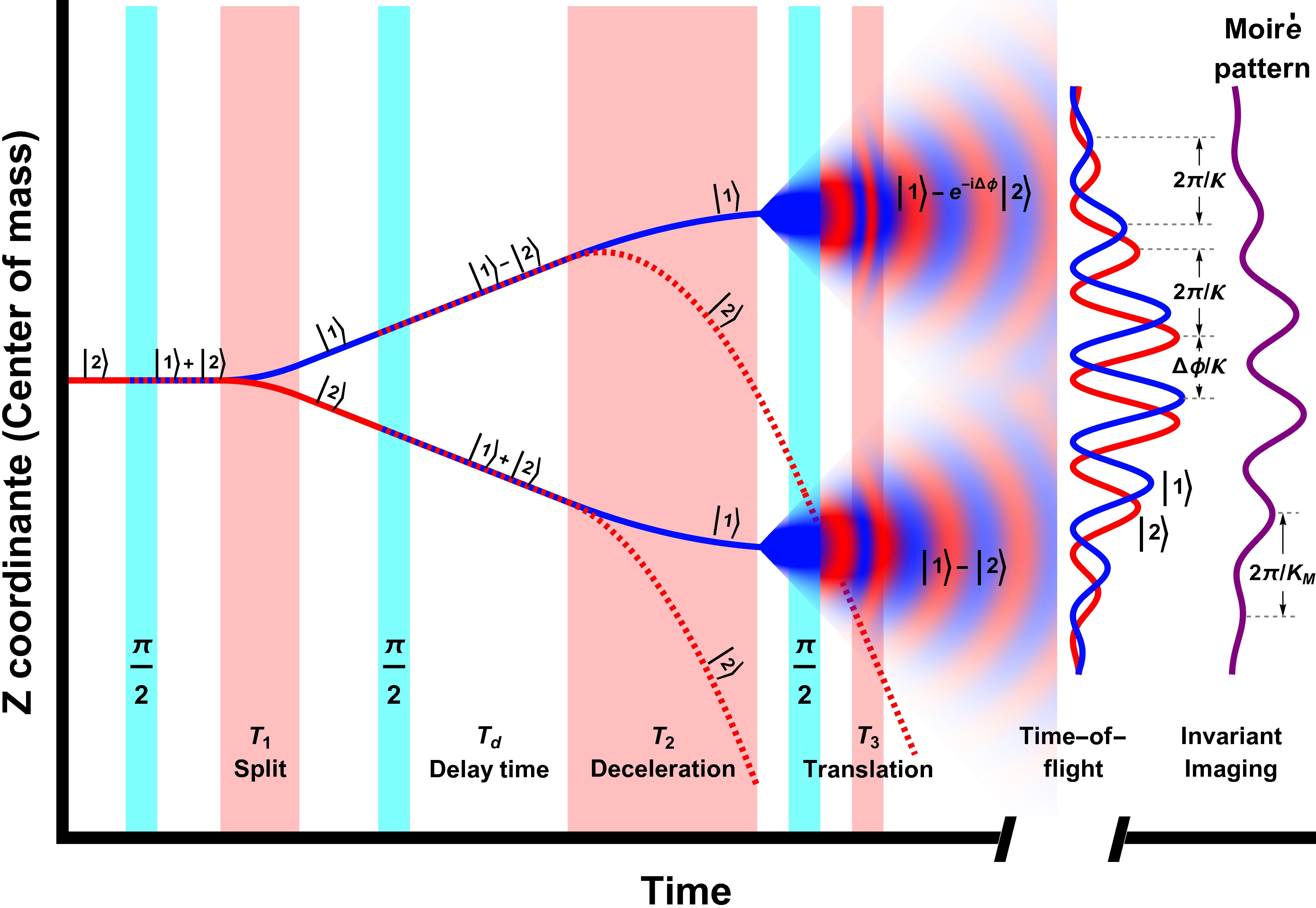}}
\caption{Experimental sequence (schematic representation, not to scale): The BEC is initialized in state $|2\ket\equiv|{F=2,m_F=2}\ket$, represented by the red line and released from the magnetic trap (not shown). After a time-of-flight of $\sim1$\,ms an RF $\pi/2$ pulse (light blue) transfers each atom into an equal superposition of $|1\ket+|2\ket$, where  $|1\ket\equiv|{F=2,m_F=1}\ket$ is represented by the blue line. The magnetic field gradient pulses are in pink. See text for more details. After time-of-flight, the two expanded wavepackets overlap and interfere with one another, forming two interference patterns, one of the $|1\ket$ state and one of the $|2\ket$. Our absorption imaging method is invariant to the internal state and so a moir\'e pattern is formed on the CCD image. Red, blue and purple, have the same meaning as in Fig.\,1(a).
\label{fig:sequence}}
\end{figure}

\begin{figure}
\centerline{
\includegraphics*[angle=0, width=\columnwidth]{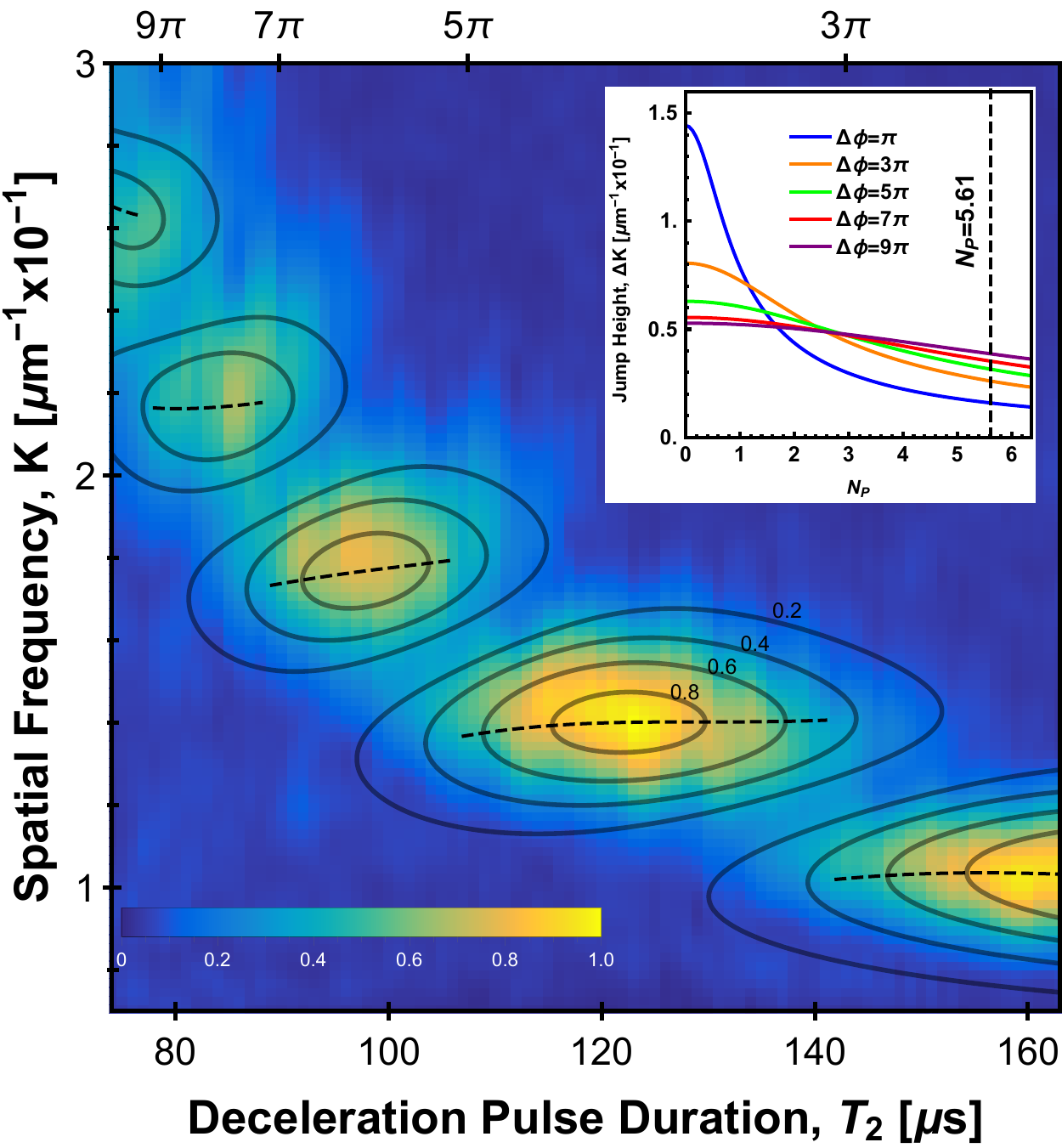}}
\caption{Comparison of the experimental data to a simplified analytical phenomenological model: The absolute value of the FT of the moir\'e pattern (AFT), as a function of the deceleration pulse duration, $T_2$, and the Fourier variable $K$. The data, represented by a heat map, is the same as in Fig.\,2. The contour map lines, representing equi-AFT lines, are the result of our phenomenological model with no free parameters (see text). The input parameters $\kappa(T_2)$, $\Delta\phi(T_2)$ and $N_p=(2/\pi)\kappa\sigma=5.6$, are independently measured (see Methods and Extended Data). The maxima of both the model and the data were normalized to $1$. In the inset, the jump height as a function of the number of periods in our finite system, whether these are potential periods as in Eq.\,1 or atomic density periods as in our experiment. Each line represents a different jump, classified by their $\Delta\phi$. The jump corresponding to $\Delta\phi=\pi$ is expected to appear at $T_2\approx300\,\mu$s, and is however below our noise limit for detection. The lines are calculated from the same model as the main figure. As explained in the text, at high $\Delta\phi$ where the two envelopes are completely separated, a universal behavior kicks in, whereby all the jumps are of the same height.
\label{fig:model}}
\end{figure}

Essential to our experiment is a short third magnetic gradient pulse of fixed duration $T_3=30\,\mu$s, determining the relative translation $\Delta z$, or the equivalent relative phase $\Delta\phi$. This is an outcome of two processes: first, the two spin states are exposed to a different magnetic potential, and second, due to the spin-dependent force, the original two Gaussian envelopes that emerged after the deceleration pulse and third $\pi/2$ RF pulse, each holding two spin wavepackets, now turn into two partially overlapping Gaussian envelopes each holding an interference pattern of one of the spins (Fig.\,3). The $T_2$ deceleration pulse has a crucial role, and consequently the data is presented as a function of $T_2$. On the one hand it determines the spatial and momentum difference between the same-spin wavepackets before they expand to form interference patterns, thus determining the final wavenumber $\kappa$ of the two constituent patterns. On the other hand, as it fixes the absolute distance of the atoms from the chip during the translation pulse, it influences the magnetic field gradient (as the gradient is not linear) and hence the differential momentum applied to the two spins, and this in turn, determines the final spatial translation $\Delta z$ and therefore the relative phase $\Delta\phi=\kappa\Delta z$.  As we vary $T_2$, we follow the experimental trajectory depicted by the red line in Fig.\,1(c).

To conclude the description of the experiment, let us note that in our experiment $\kappa\sigma $ is a constant due to a general conservation law concerning the unitary evolution of a pair of Gaussian wavepackets of the same spin in free space or in smooth potentials.  This law applies to our interferometric sequence following the splitting pulse whose  duration $T_1$ is kept constant in the experiment. The quantity $\sqrt{(\kappa\sigma)^2+(\Delta z/2\sigma)^2}$, where $\Delta z$ is the distance between the Gaussian centers, is a constant of the evolution, which is approximately equal to $\kappa\sigma$ at the time of observation. This conservation is most vividly visualized by the evolution of the Wigner function in phase space. The unitary evolution during the interferometric sequence is nothing but phase space rotation with appropriately scaled phase space coordinates~\cite{Margalit2019}, hence the shape of the Wigner function, including the number of periods $N_P=2\kappa\sigma/\pi$, is constant (see SI).

Figure\,2 presents the measured wavenumber $K_M$ of the moir\'e pattern, which is extracted from the Fourier Transform (FT) of the data, as a function of $T_2$. The value of $K_M$  exhibits a clear rigidity, and singularity points at which this value abruptly changes. The rather good agreement with the data of our numerical simulation, in which care was taken to take account of all the experimental conditions (see SI), ensures a good understanding of the experimental apparatus, but does not give insight into the effect itself. Such insight is obtained by utilizing a simplified analytical phenomenological model, as presented in Fig.\,4. Again, good agreement with the data is obtained.

The simplified analytical model on which Fig.\,4 is based, is described in the following. The model is phenomenological, as its input parameters $\kappa(T_2)$, $\Delta\phi(T_2)$ and $N_p=(2/\pi)\kappa\sigma=5.6$, are fixed by independent measurements [$\sigma(T_2)$ is also measured independently, see SI]. The $\Delta\phi$ corresponding to each jump are presented in the upper x-axis of Fig.\,4. The inset shows that the magnitude of the singularity jumps in $K_M$ for each specific $\Delta\phi$ is determined by the number of periods in the system, $N_p$. For a large number of periods the jumps disappear, at which point a smooth function of $K_M$ as a function of $T_2$ is expected.

In our simplified model we extract $K_M$ from the absolute value of the positive spatial frequency part ($K>0$) of the FT (AFT), of the sum of patterns in Eq.\,1
\begin{equation}
{\rm AFT}\equiv|{\cal F}V^{(+)}(K)|=e^{-\frac12\sigma^2(K-\kappa)^2}\left|cos\left(\frac{K}{2\kappa}\Delta\phi\right)\right|.
\label{eq:absFourier}
\end{equation}
The peak of the FT is at $K_M=\kappa$ when $\Delta\phi=2\pi n$ is an integer multiple of $2\pi$, where the cosine function is peaked at the same $K=\kappa$ as the Gaussian peak.
In general, $\Delta\phi=2\pi n+\alpha$, where $-\pi<\alpha<\pi$, and consequently the cosine peak shifts to $K=\kappa/(1+\alpha/2\pi n)$, and therefore the peak of the FT shifts to $K_M<\kappa$ or $K_M>\kappa$, depending on whether $\alpha$ is positive or negative, respectively.
The value of $K_M$ in Figs.\,1 and~\,4 is calculated by finding numerically the maximum of the FT in Eq.\,2.
Note that a similar wavenumber $K_M$ is obtained by fitting the real-space moir\'e pattern [of Eq.\,1 or the CCD images]  to the form $V(z)=A\exp[-(z-\bar{z})^2/2\bar{\sigma}^2][1+v\sin(K_Mz+\bar{\phi})]$, where $\bar{z}$, $\bar{\sigma}$ and $\bar{\phi}$ are the center position, effective width and phase of the moir\'e pattern, respectively, $v$ is the visibility and $A$ is a normalizable constant. This fitting is valid as long as the two patterns forming the moir\'e pattern significantly overlap.

An intuitive understanding of the effect can be gained by observing the variation of the phase of oscillation over the size of the moir\'e pattern, where the amplitude is dominated by the first term in Eq.\,1
at $z<z_1$ and becomes dominated by the second term at $z>z_2$ (here $z_1<z_2$).
By writing the positive spatial frequency part of Eq.\,1 in the form
\begin{equation}
V^{(+)}(z)=\frac12 e^{i\kappa z}\left[e^{i\phi_1}e^{-(z-z_1)^2/2\sigma^2} +e^{i\phi_2}e^{-(z-z_2)^2/2\sigma^2} \right],
\label{eq:V_plus}
\end{equation}
where $\phi_j=2\phi-\kappa z_j$,  we can see that the phase inside the brackets varies from about $\phi_1$ to about $\phi_2$ between the two points $z_-=\bar{z}-2\sigma^2/\Delta z$ and $z_+=\bar{z}+2\sigma^2/\Delta z$ (where $\bar{z}=\frac12(z_1+z_2)$), because at $z_-$ the amplitude of the first term is larger than that of the second term by $e^2$ so that $e^{i\phi_1}$ dominates, and vice versa for $z_+$ where the second term dominates.
Apart for the phase gradient $\kappa$ due to the term $e^{i\kappa z}$ in Eq.\,3, another phase gradient $\delta k\sim –(\Delta\phi-2\pi n)/(z_+-z_-)=-\alpha\Delta z/4\sigma^2$ appears along the moire pattern due to the term in the brackets.
This phase gradient [Fig.\,1(b)], gives rise to the observed change of $K_M$ relative to $\kappa$, and hence the jumps in $K_M$ when $\Delta\phi$ becomes closer to the next multiple of $2\pi$ with $n\to n+1$, and where $\alpha$ jumps from $+\pi$ to $-\pi$. More details are provided in the SI.

Let us also briefly discuss the possibility of having two different wavenumbers $\kappa_1$ and $\kappa_2$ for the two constituents in Eq.\,1. In fact, in our experiment, they are different by a few percent (Fig.\,S1), as the spin-$|2\rangle$ state is affected differently by the third gradient pulse relative to the spin-$|1\rangle$ state. We find that the observed rigidity is very robust to these changes. Specifically, when we input into our phenomenological model the exact experimental values for the different constituent wavenumbers, we get exactly the same plateaus (up to 0.1\%).

Finally, we note that the concept of moir\'e pattern can be extended to the realm of non-overlapping constituents by using the Fourier method presented above.
Here, each pair in an incoming flux, undergoes a joint Fourier analysis, which is sensitive to the correlation between the phases.
In particular we find that if we keep $\Delta z$ constant Eq.\,2 becomes
\begin{equation}
|{\cal F}^{(+)}V(K)|={\rm e}^{-{1\over 2}(\pi N_p/2\Delta\phi)^2(K\Delta z-\Delta\phi)^2}|\cos {1\over 2} K\Delta z|.
\end{equation}
This exhibits universal plateaus, $K_M\Delta z=2\pi n$, at intervals $\pi(2n-1)<\Delta\phi<\pi(2n+1)$ for large $\Delta\phi$, i.e. all  $\Delta\phi>N_p$ (see SI). We note that the latter condition implies $\Delta z>\sigma$, hence the two patterns are separated in space. Each pattern has a dominant periodicity $\kappa$, yet their sum exhibits, remarkably, a very different periodicity that is quantized at $K_M=2\pi n/\Delta z$. As shown in the SI, taking from an incoming flux a sub-sample of pairs with constant $\Delta z$, this quantization of $K_M$ may be used as a novel probe of the features of the emitting source.

In conclusion, we have shown both theoretically and experimentally a unique behavior of the one-dimensional moir\'e pattern. This includes a varying moir\'e wavenumber $K_M$ while the constituent wavenumbers $\kappa$ are constant, as well as rigidity and singularity jumps, whereby $K_M$ is constant while $\kappa$ and $\Delta\phi$ are changed. Our work also uncovers a broader universal 1D moir\'e phenomena in Fourier space, even when the two envelopes do not overlap in real space.
As moir\'e patterns are used in a wide range of applications\,\cite{review}, such features may prove both inspiring and useful for a variety of fundamental studies of one-dimensional periodic phenomena, as well as for technological applications.

\section*{Acknowledgments}
We are grateful to Zina Binstock for the electronics, and the BGU nano-fabrication facility for providing the high-quality chip. This work is funded in part by the Israeli Science Foundation, the DFG through the DIP program (FO 703/2-1), and the program for postdoctoral researchers of the Israeli Council for Higher Education.

{\bf Data availability} The data that support the findings of this study are available from the corresponding author upon reasonable request.

{\bf Author Contributions} O.A. and O.D. performed the experiments, analysed the data, and contributed to its theoretical interpretation. Z.Z. and Y.M. participated in the initial stages of building, running the experiment, and discussing the results. S.M. performed a preliminary theoretical analysis. Y.J. and B.H. developed the theoretical model and performed the simulations. Y.M. assisted with the theoretical interpretation. R.F. oversaw the different tasks. All authors contributed to the writing of the paper.


\section*{Methods}

\subsection{Detailed experimental scheme}
The experiment is based on $^{87}$Rb atoms cooled using a standard reflection-MOT apparatus. The atoms are then loaded into an Ioffe-Pritchard magnetic trap, and forced evaporation cooling, using a radio-frequency (RF) knife, is applied until degeneracy is reached. The BEC, containing $\sim10^4$ atoms, is released from the trap in the $\vert2\rangle\equiv|F=2,m_F=2\rangle$ state, and the experiment is conducted during free-fall. As the BEC quickly expands and becomes dilute, atom-atom interaction is negligible. The whole experiment is performed under a constant bias magnetic field of $\sim35\,$G which isolates an effective two-level system of $|2\rangle$ and $|1\rangle\equiv	|F=2,m_F=1\rangle$. The wavepacket is transferred into an equal superposition of the two internal states, $(\vert1\rangle+\vert2\rangle)/\sqrt{2}$, using an on-resonance RF $\pi/2$ pulse. The transition frequency of this two-level system is $\sim25\,$MHz. By applying a magnetic field gradient pulse of duration $T_1=4\,\mu$s, the wavepacket is split into two distinct trajectories with different momenta. The magnetic field gradient originates from a current of $I=1.1\,$A flowing along three parallel wires in alternating directions. This configuration of the wires helps reduce the phase noise originating from the chip currents \cite{Margalit2019}. \par
After the initial splitting pulse, the two wavepackets freely propagate for a time ${T_d}_1=230\,\mu$s. During this time we apply a second $\pi/2$ pulse which transfers each of the trajectories, previously in a pure state of either $|1\rangle$ or $|2\rangle$ into an equal superposition of $(|1\rangle-|2\rangle)/\sqrt{2}$ and $(|1\rangle+|2\rangle)/\sqrt{2}$, respectively. The wavepackets are then decelerated relative to one another by a second magnetic gradient field pulse of varying duration $T_2$, which causes the $|1\rangle$ components from each of the wavepackets to have roughly the same momentum. The $|2\rangle$ components are ejected from the interferometer region-of-interest and are ignored for the rest of this experiment. This decelerating is possible due to the non-linearity of the magnetic field; that is, each wavepacket in internal state $|1\rangle$ feels different acceleration since the magnetic field gradient is not constant in space. The same non-linearity also causes the wavepackets to go through a focal point and then expand at an increased rate. \par
After the deceleration pulse, we apply a third and final RF $\pi/2$ pulse, creating two superpositions of $|1\rangle$ and $|2\rangle$. In total, our quantum system now consists of two spatially separated wavepackets, each in a superposition of $|1\rangle$ and $|2\rangle$. The atom freely propagates until we apply a third magnetic field gradient pulse of duration $T_3=30\,\mu$s. The third pulse has opposite polarity compared to the first two pulses, such that the acceleration due to the magnetic field gradient is directed upward. By reversing the polarity of the third magnetic field, we can achieve slightly longer measurements since the BEC remains in the field of view of the CCD for a longer time.  This third pulse imparts a differential phase $\Delta\phi$ and a differential position $\Delta z$, on the two spin interference patterns. The timing of the third magnetic field gradient pulse is ${T_d}_2=410\,\mu$s after the start of $T_2$. \par
The bias magnetic field is shut down $660\,\mu$s after the last magnetic field gradient pulse, after which the wavepackets fall under gravity and expand for another $14\,$ms. At the end of the experimental cycle, we image the interference pattern using a standard absorption imaging technique. Since the bias magnetic field is turned off before the imaging pulse, our imaging beam is invariant to the two energy levels and can not distinguish between $|1\rangle$ and $|2\rangle$. This measurement is repeated for different values of $T_2$, and its results are presented in the main text. \par
As auxiliary measurements, we repeat the same experimental sequence described above, where we change the third RF pulse, of duration ${T_R}_3$ and Rabi frequency $\Omega$, from from $\Omega{T_R}_3=\pi/2$ to $\Omega{T_R}_3=0$ or $\Omega{T_R}_3=\pi$. Effectively, these two extra measurements, presented in Fig.\,\ref{fig:SM1}, measure the periodicity of the single-state interference pattern, $\kappa_i$.

\subsection{Data analysis}
An example of the raw data can be seen in the insets of Fig.~2. The analysis starts by fitting a Gaussian function of width $\sigma_0$ to the interference pattern. While the accuracy of this fit is limited, it gives a rough estimate of the envelope function. Using these results, we subtract the Gaussian envelope from the data and calculate the absolute value of the Fourier Transform (FT) along the elongated axis of the interference pattern ($z$, gravity). We cut the spatial frequencies lower than $1/0.9\sigma_0$. This removes the central peak of the FT. While there is some redundancy in filtering the central peak (low-frequency cutoff and subtraction of the envelope from the data), we found this method to give more consistent results. We now locate the two highest maxima of the FT spectrum and calculate their relative intensity. We take the highest value as the main peak, $K_M$, and the second peak is recorded only if its relative intensity is at least $20\%$. These values are plotted in Fig.~2(a). \par
Going back to the original data in real space, we now perform a second fit. The function used is a Sine function multiplied by a Gaussian envelope
\begin{equation}
f_1(z)=A \eexp{-(z-\bar{z})^2/2\bar{\sigma}^2}\left[ 1+v{\rm sin}\left(K_M z+\bar{\phi}\right)\right]+c,
\label{eq:fitrealspace}
\end{equation}
where $A,\,\bar{z},\,\bar{\sigma},\,v,\,\bar{\phi},$ and $c$ are the fitting parameters. In this fit, we set the value of $K_M$ to be the main peak of the FT. From these results, we take $v$ to be the visibility of the interference pattern, plotted in Fig.~2(b). \par
For our phenomenological model we fit the visibility, Fig.~2(b), to the function
\begin{equation}
v(T_2)=\frac{v_0}{2}{\rm cos}\left(\frac{a}{T_2^2}+\phi_0\right)+c,
\label{eq:fitvisibility}
\end{equation}
where $c,\,v_0,\,a,$ and $\phi_0$ are the fitting parameters. From the results of this fit we obtain the differential phase, given by
\begin{equation}
\Delta\phi(T_2)=\frac{a}{T_2^2}+\phi_0,
\label{eq:diffphase}
\end{equation}
where $a=(163\pm2)\cdot10^3\,\mu{\rm s}^2$ and $\phi_0=1.3\pm0.2$ Rad are the result of the fit to $v(T_2)$. We also fit each of the single-state wavenumbers, plotted in Fig.\,\ref{fig:SM1}, to the function
\begin{equation}
\kappa_i(T_2)=\frac{2\pi a_i}{\sqrt{T_2+b_i}},
\label{eq:fitvisibility}
\end{equation}
where $a_i$ and $b_i$ are the fitting parameters. The results of these fits are $a_1=0.175\pm0.004\,\mu{\rm m}^{-1}\cdot\mu{\rm s}^{1/2}$ and $b_1=-56\pm1\,\mu$s for the case of $\Omega{T_R}_3=0$ and $a_2=0.183\pm0.004\,\mu{\rm m}^{-1}\cdot\mu{\rm s}^{1/2}$ and $b_1=-56\pm1\,\mu$s for the case of $\Omega{T_R}_3=\pi$. \par
For our simplified model, presented in the main text, we use $\kappa(T_2)=\frac{1}{2}\left[\kappa_1(T_2)+\kappa_2(T_2)\right]$. For the independent measurement of $N_P$, presented in Fig.~\ref{NPplot}, we use the aforementioned values for $\kappa_1(T_2)$ and $\kappa_2(T_2)$ together with the values of $\bar \sigma(T_2)$ extracted from the fit to Eq.~(\ref{eq:fitrealspace}). We then calculate $N_P(T_2)=(2/\pi)\kappa(T_2)\bar\sigma(T_2)$, and average the results to get $N_P=5.61\pm0.03$.


\section*{Supplementary Information}

\section{Model and conservation laws}
\label{sec:conservation}

In this section we derive the basic properties of of our model following the experimental scenario. We show that the interference patterns has the form of Eq. (1) of the main text with $\theta_1=\theta_2$, explain the phase relation $\Delta\phi=\kappa\Delta z$ and prove  the conservation law for $\kappa\sigma$. This derivation is based on a general argument that allows for interactions in the initial state and also allows for arbitrary form of the magnetic field $B_y(z)$.

Consider an initial time $t_0$ after the second $\pi/2$ pulse and before the deceleration pulse of duration $T_2$ (see Fig. 3, main text).
We consider only states $|1\rangle$ ignoring states $|2\rangle$ (whose separation from the states $|1\rangle$  is achieved later with the deceleration pulse), thus we have two states $|1\rangle$ with different momenta and we follow their unitary evolution in time. We define the overlap of these states $\psi_a(z,t_0),\, \psi_b(z,t_0)$ as (using $\int_z=\int_{-\infty}^\infty dz$)
\begin{equation}\label{t0}
\int_z\psi^*_a(z,t_0)\psi_b(z,t_0)=\eexp{-\half \Gamma^2+i\chi}
\end{equation}
with $\Gamma,\,\chi$ real
and the initial states normalized as $\int_z|\psi_a(z,t)|^2=\int_z|\psi_b(z,t)|^2=1$.
In our experiment the initial time $t_0$  is after the splitting pulse and quite shortly after the BEC is released from the trap, so that the wave functions include effects of interactions and may deviate considerably from a Gaussian form.

We next consider the full evolution with arbitrary
magnetic fields during $T_2$ and $T_3$,
however  we follow two evolution scenarios: one for states that stayed in the spin state $|1\rangle$ during the third $\pi/2$ pulse and one for the states that flipped into the spin state $|2\rangle$ during this RF pulse.
The superposition of the two scenarios corresponds to the wave function
\begin{equation}
\frac{1}{\sqrt{2}}(\psi_{1a}+\psi_{1b})|1\rangle+\frac{1}{\sqrt{2}}[\psi_{2a}+\psi_{2b})|2\rangle,
\end{equation}
keeping the normalization of
The initial states $(\psi_a(z,t)+ \psi_b(z,t))|1\rangle$.
The pair $\psi_{1a},\psi_{1b}$ forms the interference pattern of spin 1 while the pair $\psi_{2a},\psi_{2b}$ forms that of spin 2, the summation of both interferences gives the observed moir\'e pattern.
For either scenario
the time evolution represented by the evolution operator
$U_i(t,t_0,z)$, is identical for $\psi_{ia}(z),\psi_{ib}(z)$ (common Hamiltonian for a given spin state $i=1,2$), hence the overlap at the final time $t$ is
\begin{eqnarray}\label{evolution}
&&\int_z \psi^*_{ia}(z,t)\psi_{ib}(z,t)= \int_z \psi^*_{a}(z,t_0)U_i^\dagger(t,t_0,z)\times\nonumber\\&&U_i(t,t_0,z)\psi_{b}(z,t_0)
=\int_z \psi_{a}^* (z,t_0) \psi_{b}(z,t_0)
\end{eqnarray}

\begin{figure}[h]
\centerline{
\includegraphics[trim={0cm 0 0cm 0cm},clip,width=\columnwidth]{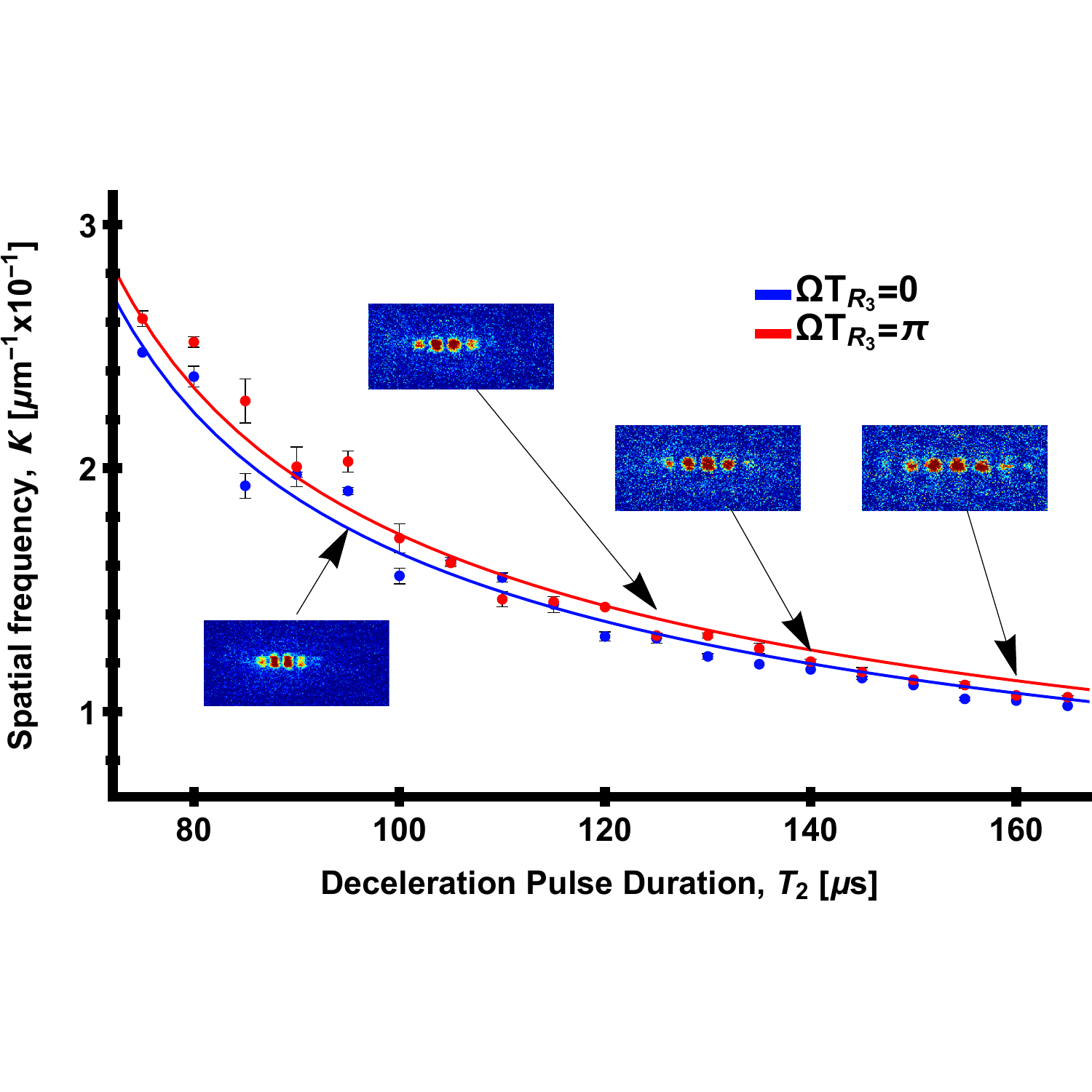}}
\caption[]{Single state interference pattern periodicity, $\kappa$. The experimental sequence presented in Fig.~3 is repeated twice, once without the last RF pulse, $\Omega{T_R}_3=0$ (green) and a second time with the last RF pulse twice as long $\Omega{T_R}_3=\pi$ (purple). From these measurements we get the single-state interference pattern. Error bars are calculated from the SEM. The solid lines are a phenomenological fit.}
\label{fig:SM1}
\end{figure}

\begin{figure} \centering
\includegraphics [width=.45\textwidth]{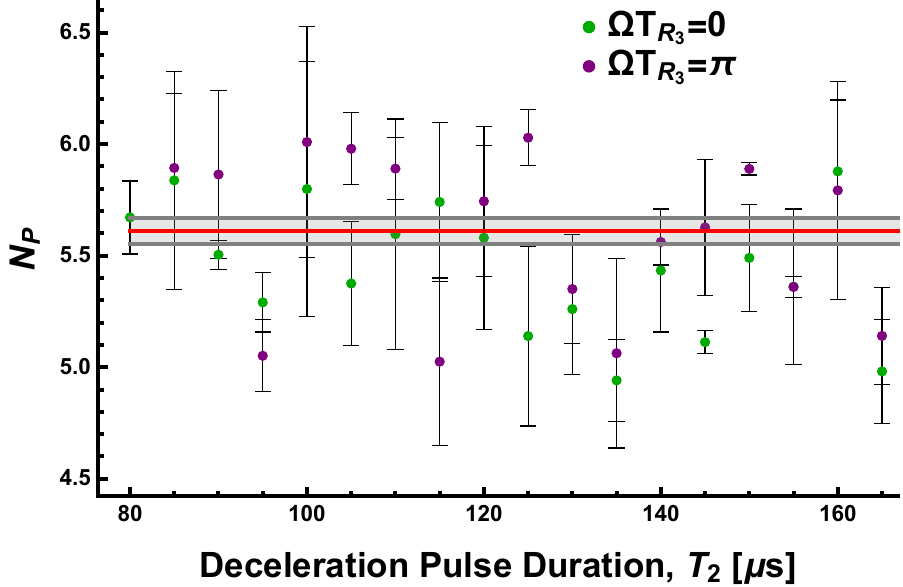}
\caption {The number of observable peaks $N_P=\frac{2}{\pi}\kappa\sigma$ for the interference of a single spin state, either 1 ($T_{R_3}=0$) or 2 ($T_{R_3}=\pi$). }
\label{NPplot}
\end{figure}

From this conservation law of the overlap integral we derive more explicit conservation laws that involve parameters of the wavepackets. We next assume that the wave-packets after the gradient pulses are Gaussians with equal time-dependent width $\sigma_i(t)$
For the pair of wavepackets with the same spin,
but different center positions and momenta. In our experiment the final wavepackets are fairly close to a Gaussian and the two spin states have similar widths, see Figs.\,\ref{fig:SM1},\ref{NPplot}. This is indeed expected as long as the two wavepackets, which originate from the same initial wave-packet before splitting, propagate in free space and in a potential that has a constant curvature $\partial^2 V/\partial z^2$ over the range occupied by the two wavepackets.
This condition is satisfied in the experiment since the potential due to the magnetic field during the gradient pulses is generated by current wires whose distance from the atoms (about 100\,$\mu$m) is much larger than the distance between the wavepackets or their width (a few $\mu$m).

If this condition is satisfied then the interference terms have the form
\begin{equation}
\psi^*_{ia}(z,t)\psi_{ib}(z,t)=A_i^2 e^{-[(z-z_{ia})^2+(z-z_{ib})^2]/4\sigma_i^2}e^{-i\kappa_i z+i\phi_i},
\label{eq:interf_term}
\end{equation}
where $A_i=(2\pi\sigma_i^2)^{-1/4}$ is the normalization constant and $z_{ia},z_{ib}$ are the corresponding Gaussian centers. The explicit forms of $\kappa_i$ and $\phi_i$ are given below, though for now they are not needed.

We now use $(z-z_{ia})^2+(z-z_{ib})^2=2(z-z_i)^2+\half(\delta z_i)^2$, where $\delta z_i=z_{ib}-z_{ia}$ and
$z_i=\frac12(z_{ia}+z_{ib})$ is the average position of each pair of states.
The interference pattern from \eqref{eq:interf_term} is then
\begin{equation}\label{eq:pattern}
\re [\psi^*_{ia}(z,t)\psi_{ib}(z,t)]=A'_i\eexp{-(z-z_i)^2/2\sigma_i^2}\cos[\kappa_i z-\phi_i]
\end{equation}
where $A'_i=A_i^2\eexp{-(\delta z_i)^2/8\sigma_i^2}$.

 Integrating Eq.~(\ref{eq:interf_term}) yields
\begin{equation}
\int_z \psi_{ia}^*(z,t)\psi_{ib}(z,t)=e^{-\frac12 [(\delta z_i/2\sigma_i)^2+(\kappa_i\sigma_i)^2]}
e^{-i\kappa_iz_i+i\phi_i} .
\end{equation}
By comparing this form to Eq.~\eqref{t0} the conservation law for the overlap integral yields our central result
\begin{eqnarray}
\Gamma^2 =&& \left(\frac{\delta z_i}{2\sigma_i}\right)^2+(\kappa_i\sigma_i)^2, \label{eq:Gamma}\\
\chi=&&-\kappa_iz_i+\phi_i  \label{eq:chi1}
\label{eq:chi}
\end{eqnarray}\\
Note that all the parameters $\delta z_i, \sigma_i, \kappa_i, z_i,\phi_i$ are time dependent so that the right hand sides of Eqs. (\ref{eq:Gamma},\ref{eq:chi1}) are conserved in time for either spin state and are in fact spin independent, since $\Gamma,\,\chi$ are common to both spins.

Eq. \eqref{eq:pattern} can now be identified with the interference part of each of the two patterns in Eq. (1) in the main text, with $\phi_i=\kappa_i z_i-2\theta_i$. By comparing this phase term with Eq.~(\ref{eq:chi}) we identify $\theta_i=-\chi/2$. As $\chi$ is a conserved quantity that is common to both spins we therefore conclude that $\theta_1=\theta_2$. This is an exact result exhibiting the quantum signature of our implementation of Eq. (1) in the main text as it follows the quantum evolution from a common source.

Note that $\sigma_i$ and $\delta z_i$ are nearly the same for $i=1,2$, determined mainly by the size, separation and relative momenta of the common parent state entering the third $\pi/2$ pulse.
While the curvature of the third gradient pulse of duration $T_3$ affects the sizes and relative distances and  momenta within pairs of states differently for each spin, this effect is quite minor due to the fact that $T_3$ is much shorter than $T_2$.
As $\delta z_1\approx \delta z_2$ and $\sigma_1\approx\sigma_2$, Eq.~(\ref{eq:Gamma}) iimplies that  $\kappa_1\approx \kappa_2$, as observed in the experiment (see Methods).

In general, the phase difference between the two patterns $\Delta\phi= (\kappa_1z-\phi_1)-(\kappa_2z-\phi_2)$  is, in general $z$-dependent. However, when we use the approximation $\kappa_1=\kappa_2=\kappa$ this phase difference becomes
\begin{equation}
\Delta \phi=\kappa(z_2-z_1)\equiv \kappa\Delta z.
\end{equation}

To conclude that $\kappa\sigma$ is independent of $T_2$ we need to show that $\frac{\delta z }{2\sigma}$ at the time of observation is relatively small.  In fact, in order to observe a well defined interference one needs the separation $\delta z$ to be smaller than the combined width of the wavepackets, thus reliable experimental data must have $\delta z\lesssim 2\sigma$.
In Fig. \ref{NPplot} we present the product $N_P=\frac{2}{\pi}\kappa\sigma$, found by fitting our real space data. The data is therefore consistent with the conservation law and determines $\kappa\sigma=\frac{\pi}{2}5.61=8.81$. The 1$^{st}$ term of Eq. \eqref{eq:Gamma} is then $\lesssim (\frac{1}{8.81})^2\approx 1\% $ of the second term and can indeed be neglected.

Let us now derive the form of the interference Eq.~\eqref{eq:interf_term} and identify explicit forms for the interference wavenumber $\kappa(t)$ and for the width $\sigma(t)$ during the evolution in free space. The general shape of two Gaussian wave functions with the same widths at any time is (the following applies to either spin state and we ignore the index $i=1,2$, for simplicity)
\begin{eqnarray}
\psi_a(z,t) &=& Ae^{-(z-z_a)^2/4\sigma^2+\half i\alpha(z-z_a)^2+ik_a(z-z_a)+i\phi_a}, \nonumber \\
\psi_b(z,t) &=& Ae^{-(z-z_b)^2/4\sigma^2+\half i\alpha(z-z_b)^2+ik_b(z-z_b)+i\phi_b},
\label{eq:psiab}
\end{eqnarray}
where $\alpha(t)$ is a coefficient of the quadratic phase that evolves when the Gaussian expands or shrinks and is the same for the two wavepackets if the width $\sigma(t)$ is common. Here also the centers $z_a,\,z_b$, the phases $\phi_a,\,\phi_b$ and $A=(2\pi\sigma^2(t))^{-1/4}$ are time dependent;
 $\hbar k_a,\hbar k_b$ are (time independent) momenta of the wavepackets.

For a particle with mass $m$ the free space evolution of the Gaussian width is
\begin{equation}\label{eq:sigma}
\sigma(t)=\sqrt{\sigma_m^2+(\hbar/2m\sigma_m)^2(t-t_m)^2},
\end{equation}
where $\sigma_m$ is a minimum size occurring at time $t=t_m$ in the past or future history of the evolution if it occurs in free space. In our case this minimal wavepacket size occurs at some time (focusing time) after the end of the deceleration pulse (usually after the translation pulse $T_3$).
The quadratic phase coefficient is given by
\begin{equation}
\alpha(t)=\frac{m}{\hbar}\frac{\dot{\sigma}}{\sigma}=\frac{\hbar(t-t_m)/4m\sigma_m^4}{1+(\hbar/2m\sigma_m^2)^2(t-t_m)^2},
\label{eq:alpha}
\end{equation}
so that $\alpha=0$ at the focusing time $t=t_m$ and $\alpha(t)\to m/\hbar (t-t_m)$ after a long time where $\sigma(t)\gg \sigma_m$.
The phase of the interference term of $\psi_a$ and $\psi_b$ of Eq.~(\ref{eq:psiab}) can be written as
\begin{equation}
-(\alpha\delta z-\delta k)(z-\bar{z})+\delta\phi-\bar{k}\delta z,
\end{equation}
where $\bar{z}=\frac12(z_a+z_b)$, $\bar{k}=\frac12(k_a+k_b)$ and $\delta k=k_b-k_a$ are the average momentum and momentum difference between the two wavepackets, while $\delta\phi=\phi_b-\phi_a$.
We therefore identify
\begin{eqnarray}\label{eq:kappa}
\kappa&=& \alpha\delta z-\delta k
\end{eqnarray}

Finally we derive an interesting relation between the properties of the state at the observation time $t_f$ of our experiment and at the focusing time $t_m$, where $t_f$ occurs at a time $T_f$ (time of flight) after $t_m$, i.e. $t_f=t_m+T_f$. The width from Eq. \eqref{eq:sigma}  and wavelength from Eqs. (\ref{eq:alpha},\ref{eq:kappa}) are given by
\begin{eqnarray} \label{eq:sigma1}
\sigma(t_f)&=&\frac{\hbar T_f}{2m\sigma_m}\sqrt{1+\xi^2}  \\
\kappa(t_f)&=&\frac{md}{\hbar T_f}\frac{1}{1+\xi^2}-\delta k \frac{\xi^2}{1+\xi^2},
\label{eq:kappa1}
\end{eqnarray}
where $d=\delta z(t_m)$ is the distance (positive or negative) between the wavepacket centers of the same spin at the focusing time and $\xi=2m\sigma_m^2/\hbar T_f$.
In our experiment the wavepacket that has a larger momentum at $t_0$ is always further away from the chip so we can choose the indices $a$ and $b$ such that always $z_b>z_a$ and $d>0$ and hance $\kappa(t_f)>0$ in the limit $\xi\ll 1$.

When the Gaussian wavepackets are much larger than their minimal size $\sigma_m$ the factor $\xi$ becomes negligible and the equations for $\kappa$ and $\sigma$ simplify considerably. This situation indeed applies to the conditions of our experiment, as predicted by our numerical simulation (see Sec.~\ref{sec:numerical}), but let us show this here by using only direct experimental evidence.
Without the focusing effect due to the positive curvature of the magnetic field potential applied during the pulses the expansion of the BEC was measured to obey the free expansion rule $\sigma(t)=\sigma(0)\sqrt{1+\omega^2 t^2}$ for any time $t$ after trap release, where $\omega\approx 2\pi\times 120\,$Hz is the trap frequenciy along $z$ in our experiment.  For $\sigma(0)=1.2\,\mu$m~\cite{Margalit2018}, free expansion in our experiment would lead after a time $t\approx 16$\,ms (from trap release to observation) to a final size $\sigma\sim \sigma(0) \omega t \approx 15\,\mu$m.
However, for the observed range of $T_2$ the measured width in our experiment is in the range  $\sigma(t_f)=30-80\,\mu$m, which is at least twice as large as the expected width in free expansion. This enlarged wavepacket size indicates that the magnetic fields focus the wavepackets to a minimal size $\sigma_m$ that is much smaller than 15$\,\mu$m and this focusing causes the enhanced expansion.  As a first step let us only assume $\sigma_m<15\,\mu$m. This puts an upper bound to $\xi$, as [using Eq.~\eqref{eq:sigma1}]:  $\xi=1/\sqrt{(\sigma(t_f)/\sigma_m)^2-1}<1/\sqrt{(2^2-1}=1/\sqrt{3}$. This bound allows us to set an upper bound for $\sigma_m=\sqrt{\hbar T_f\xi/2m}<1.7\,\mu$m (using $T_f<14$\,ms and the mass of rubidium $m=1.44\cdot 10^{-25}$\,kg). Once this bound on $\sigma_m$ is set, we can estimate that $\xi\approx \sigma_m/\sigma(t_f)<1.7/30\ll 1$ and therefore we can take the limit $\xi\to 0$ in Eqs.~\eqref{eq:sigma1} and~\eqref{eq:kappa1}, giving rise to the well known relation $\kappa=md/\hbar T_f$.
We then obtain by multiplying the two equations
\begin{equation}\label{eq:approx}
\frac{d}{2\sigma_m}\approx \kappa(t_f)\sigma(t_f)\approx \Gamma.
\end{equation}
This result can also be obtained by applying the conservation law Eq. \eqref{eq:Gamma} at $t_m$, where $\kappa(t_m)=-\delta k=k_a-k_b$, i.e.
\begin{equation}
\frac{d^2}{4\sigma_m^2}+(k_a-k_b)^2\sigma_m^2=\Gamma^2
\end{equation}
As $\delta z(t_f)=d+ \hbar\delta k T_f/m$ and $\sigma_m\approx  \hbar T_f/2m\sigma(t_f)$ we obtain $|\delta k\sigma_m|\approx |\delta z(t_f)-\delta z(t_m)|/2\sigma(t_f)\lesssim 1\ll \Gamma$, as (where $|d|\ll \sigma(t_f)$) we concluded above, hence Eq.~\eqref{eq:approx} is proved.

Having established the relations above, we can obtain estimates of the parameters of the state at the focusing time by assuming that the focusing time $t_m$ relative to the end of the translation pulse of duration $T_3$ is much smaller than the remaining time of flight $T_f$, so that $T_f\sim 14$\,ms. It then follows from $\kappa$ being in the range 0.3-0.1$\,\mu$m$^{-1}$ that $d\approx \hbar\kappa T_f/m$ is in the range 3$\,\mu$m to 1\,$\mu$m, while $\sigma_m$ is in the range 0.17-0.06$\mu$m. It then follows that $\xi$ is in the range 0.006-0.0007.
These values are reproduced in our numerical simulation (see Sec.~\ref{sec:numerical} and Fig.~\ref{fig:wppars}).

In section~\ref{sec:Wigner} below we present an additional demonstration of the conservation laws, which does not rely on the overlap integral between the wavepackets and uses, instead, arguments based on the evolution of the phase space distribution (Wigner function) of a superposition of two wavepackets.

We have shown in this section that an internally coherent system, which we may define as a system having constituents from a single coherent parent that is split into a superposition of two internal states, satisfies $\Delta\theta=0$, as each of the constituents preserve the phase $\chi$ of the parent state.
While this proof is based on properties of quantum superpositions, the same property may also appear in classical electromagnetic pulses (e.g. radio-frequency or microwave). When a single pulse with a measurable real field $\psi(x,t)=\exp[-\frac{1}{4\sigma^2}(x-vt)^2]\cos[k(x-vt)+\theta]$, representing an electric or magnetic field, is split into a pair of pulses propagating in different trajectories and then recombined with a time delay $\Delta T$, the resulting field is a superposition $\psi(x,t)+\psi(x,t-\Delta T)$.
This field has a form similar to Eq.~(1) of the main text with $\Delta\theta=0$.
Whether such an electromagnetic field is generated by a splitting and recombination process or by an engineered electronic pulse generator, it will conserve the property $\Delta\theta=0$ when it propagates through any homogeneous medium.
However, if the delay between the two pulses is generated in a dispersive medium where the group velocity $v_g$ is not equal to the phase velocity $ v_{\phi}$, then the time delay is different for the Gaussian envelope of the pulse and for the phase of the oscillations within this envelope and therefore $\theta$ is not conserved and becomes different for the two pulses. Our model may also be utilised to treat this $\Delta\theta=$constant case (see Sec.\,II).

For an example of $\Delta\theta=$constant from the optical range or the microwave range, consider a pulsed laser (or maser) whose output is a consequence of multiple reflections inside the laser cavity, each output pulse is followed by another pulse that is generated by propagating along another round trip in the laser medium, which is a dispersive medium. While the Gaussian envelope is delayed by $\Delta T_g=2L/v_g$, where $L$ is the length of the medium, the oscillations inside the envelope are delayed by $\Delta T_{\phi}=2L/v_{\phi}$ so that $\Delta\theta=\omega(\Delta T_{\phi}-\Delta T_g)$.
In a mode-locked laser $\Delta\theta$ can be locked on a certain value that is not changed when parameters of the laser are changed and hence such a source of pulses can also be made to satisfy $\Delta\theta=$constant. Some mode-locked lasers even have as their output $\Delta\theta=0$ by using a pulse-picker. Obviously, our model may be tested at the high frequencies of visible light, only once detectors are engineered with enough bandwidth so that they may follow the oscillating electric field.

\section{Fourier analysis of the Model}

In this section we identify the periodicity wavenumber $K_M$ of the moir\'e pattern modeled by a sum of two translated localized periodic patterns with a Gaussian profile [Eq.~(1) of the main text]. We consider the Fourier transform (FT) of this pattern with Fourier variable $K$.
The FT contains two terms $\propto \eexp{-\half\sigma^2(K\pm\kappa)^2}$, we keep the term that is peaked at $K>0$, up to a prefactor $\sigma\sqrt{2\pi}$,
\begin{eqnarray}
&&{\cal F}V^{(+)}(K)=\half\eexp{-\half\sigma^2(K-\kappa)^2}
[\eexp{i(K-\kappa)z_1+i\phi_1}+\eexp{i(K-\kappa)z_2+i\phi_2}]\nonumber\\
&&|{\cal F}V^{(+)}(K)|=\eexp{-\half\sigma^2(K-\kappa)^2}|\cos\half[(K-\kappa)\Delta z+\Delta\phi]|
\end{eqnarray}
and using $\Delta\phi=\kappa\Delta z$ yields Eq. (2) of the main text.
 The position at the maximum of $|{\cal F}V^{(+)}(K)|$ is defined as $K_M$ and is given by
 \begin{equation}\label{KM}
 K_M=\kappa-\frac{\Delta\phi}{2\kappa\sigma^2}\tan(\frac{K_M}{2\kappa}\Delta\phi)
 \end{equation}
Only when the phase difference $\Delta\phi$ is an integer multiple of $2\pi$, $\Delta\phi=2\pi n$, Eq.~\eqref{KM} has the trivial solution $K_M=\kappa$. Otherwise the period of the moire pattern is different than the period of its constituents: the fundamental result of this work.

\begin{figure}
\includegraphics[width=\columnwidth]{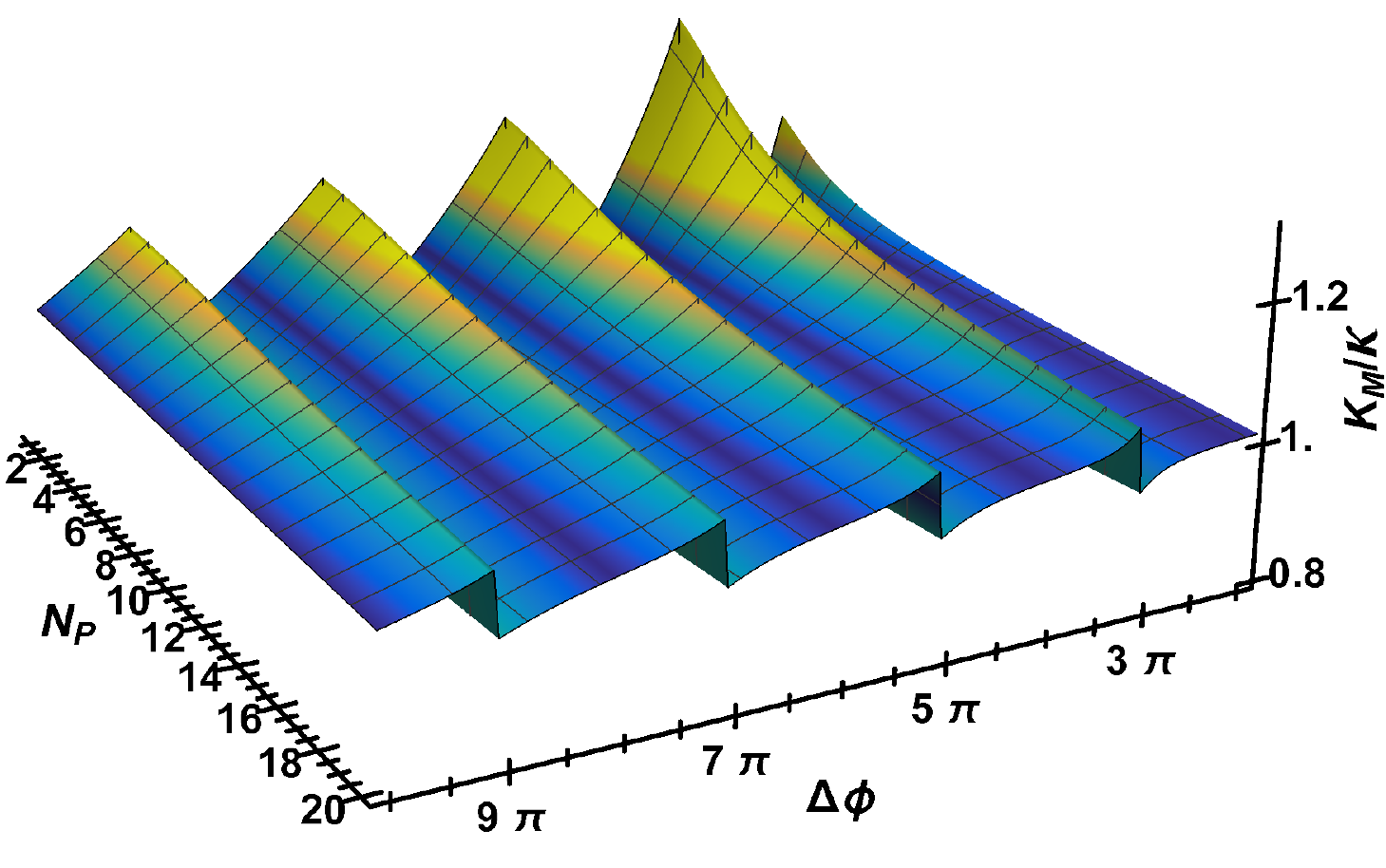}
\caption{moire pattern wavenumber $K_M$ in units of the constituent wavenumber $\kappa$ as a function of the phase difference $\Delta\phi$ and the number of periods within the system size $N_P$.}
\label{fig:K_M_phi_N_P}
\end{figure}

In Fig.~\ref{fig:K_M_phi_N_P} we present the solution for $K_M/\kappa$ at the peak of the FT in Eq. \eqref{KM} as a function of $\Delta\phi$ and $N_P$ where
\begin{equation}
N_P=\frac{2}{\pi}\kappa\sigma=\frac{4\sigma}{\lambda}
\label{eq:N_P}
\end{equation}
is the number of periods of each of the constituent patterns over the range where their envelopes are larger than $1/e^2$ of their maxima.
The value of $N_P$, as for $\kappa\sigma$, is conserved, i.e. independent of $T_2$ in our experimental setup.
The deviation of $K_M$ from $\kappa$ is
larger when the number of periods within the system size is small, while it diminishes with $N_P$ so that in the limit of an infinite periodic system, $N_P\to \infty$, the period $K_M$ of the moire pattern is just the period of the constituent patterns $\kappa$.

\subsection{Jumps}
 $K_M$ has in particular two degenerate solution at $\Delta\phi=\pi(2n+1)$ at $K_M=\kappa\pm\half\Delta K_M$ where the jump $\Delta K_M$ satisfies
\begin{equation}
\Delta K_M=\frac{\pi(2n+1)}{\sigma^2 \kappa}\cot\frac{\Delta K_M \pi(2n+1)}{4\kappa}
\end{equation}
with solutions shown in the inset to Fig. 4 of the main text.

\subsection{Rigidity}
We can gain some insight into the formation of plateaus (rigidity of $K_M$ between jumps) by looking analytically at the derivative of Eq.\,\eqref{KM} with respect to $\Delta\phi$ in between the jumps, i.e. at $\Delta\phi=2\pi n$. A plateau requires that this derivative vanishes,
\begin{equation}\label{slope}
\left. \frac{dK_M}{d\Delta\phi}\right|_{\Delta\phi=2\pi n}=\frac{\partial\kappa}{\partial\Delta\phi}-\frac{2\pi n\kappa}{(2\pi n)^2+\pi^2N_P^2}=0.
\end{equation}
This requirement can be satisfied for each $n$  if
\begin{equation}
\kappa(\Delta\phi)=\kappa_0\sqrt{\Delta\phi^2+\pi^2 N_P^2},
\end{equation}
where $\kappa_0$ is a constant.
By using the relations $\Delta\phi=\kappa\Delta z$ and $N_P=2\kappa\sigma/\pi$ Eq. \eqref{slope} is satisfied if
\begin{equation}
\sqrt{(2\sigma)^2+ \Delta z^2}=\frac{1}{\kappa_0}={\rm const.}
\end{equation}

\begin{figure}
\includegraphics[width=\columnwidth]{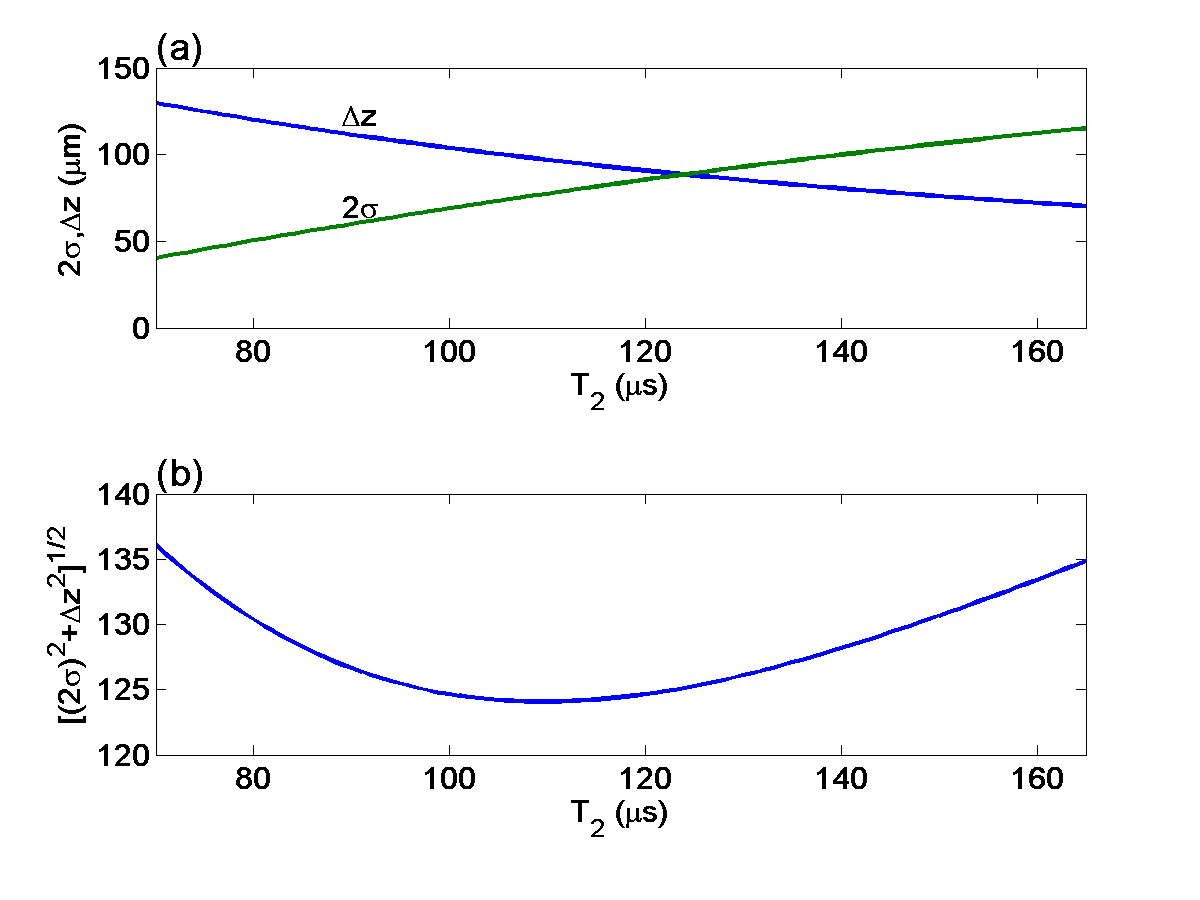}
\caption{Mutual cancellation of the variation of system size ($2\sigma$) and relative translation $\Delta z$
leading to the rigidity of moir\'e pattern periodicity in the experiment: numerical results (see Sec.~\ref{sec:numerical} below). (a) Average wave-packet size $\sigma$ increases with $T_2$ due to increased focusing strength and a resulting increased expansion speed during time-of-flight, while the translation distance decreases with $T_2$ due to a weaker final gradient pulse given further away from the chip. (b) the root sum of squares $\sqrt{(2\sigma)^2+\Delta z^2}$ varies relatively slightly over the range shown, giving rise to a minor deviation from zero slope of $K_M$ as a function of $T_2$ at the center of each of the plateaus.}
\label{fig:dzsigma}
\end{figure}

In Fig.~\ref{fig:dzsigma} we show  the dependence of $\Delta z$ and $2\sigma$ on the deceleration time $T_2$ over the range shown in Figs. 2 and~4 of the main text. The reason that $\Delta z$ depends on $T_2$ is that with a longer $T_2$ the atoms spend more time before reaching $T_3$ and also acquire a larger momentum, hence at $T_3$ the atoms are further away from the chip and experience a smaller gradient.
On the other hand the mean wavepacket width $\sigma$ increases with $T_2$ because the focusing strength and hence the speed of expansion is larger for longer $T_2$. The value of $\sqrt{(2\sigma)^2+\Delta z^2}$ in this range has therefore a relatively low variation (standard deviation of 2.4\% of the mean). This explains why $K_M$ shows rigidity over the range of $T_2$ variation.

\begin{figure}[tbh]
\centerline{\includegraphics*[width=80mm]{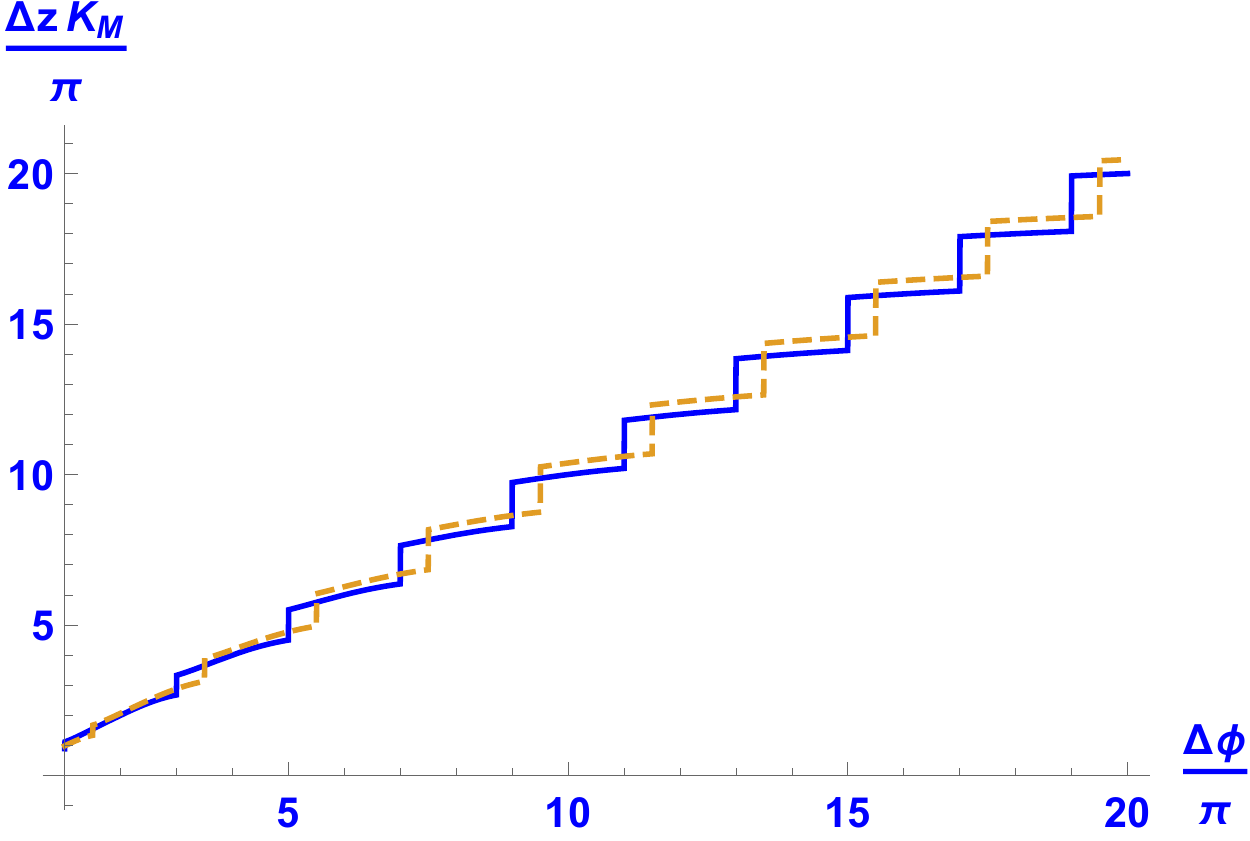}}
\caption[]{Universal plateaus in $K_M\Delta z$ for large $\Delta\phi$ when
$\Delta z$ is held constant, using $N_P=5.61$. In blue $\Delta\theta=0$ and in orange (dashed) $\Delta\theta=\pi/4$. This shows that the quantized distribution of $K_M$ is a good probe for both cases, with the possibility of differentiation by the positions of the jumps and by the values of the plateaus.}
\label{fig:universal}
\end{figure}

We consider next a scenario where $\Delta z$ is kept constant. In case of an incoming flux with random $\Delta z$, a sub-sample with constant $\Delta z$ may be used. In other cases (e.g. mode-locked laser) a constant $\Delta z$ is a feature of the source. In our experimental simulation, a constant $\Delta z$ (independent of $T_2$) may be realized by adjusting delay times between the pulses. In such a case, Eq. \eqref{KM} is written as
\begin{equation}
K_M\Delta z=\Delta\phi -\frac{(\Delta\phi)^2}{2(\kappa\sigma)^2}\tan(\half K_M\Delta z)
\label{KMfixedz}
\end{equation}
This equation is plotted in Fig.\,\ref{fig:universal}, clearly there are plateaus at large $\Delta\phi$. These plateaus become visible at $\Delta\phi\gtrsim N_P$, equivalent to $\Delta z\gtrsim \sigma$. This implies that the two patterns are separated in real space and a moir\'e pattern, in the conventional sense, is not visible. However, $\Delta\phi$ and the moir\'e wavevector are well defined. We note in particular that the plateaus are at universal values $K_M\Delta z=2\pi n$ where $n$ is defined at the center of the plateau as $\Delta\phi=2\pi n$.

We also plot in Fig.\,\ref{fig:universal} a case with a finite $\Delta\theta=\theta_2-\theta_1$ that results in shifting the argument of the $\tan$ function in Eq. \eqref{KMfixedz} by $-\Delta\theta$. The orange-dashed line corresponds to $\Delta\theta=\pi/4$, showing a similar structure except that the jump positions and plateau values are shifted.

Let us conclude. In our experiment the rigidity, namely insensitivity of $K_M$ to the change of the periodicity $\kappa$ of the constituent patterns, is achieved due to the following properties of the model:
\begin{itemize}
\item $\theta_1=\theta_2$.
\item $\kappa\sigma=\frac12\pi N_P$ constant.
\item $(2\sigma)^2+(\Delta z)^2$ constant.
\end{itemize}
While the first two conditions are exact constraints that follow from fundamental properties of the system, as shown in section\,I, the third condition is an approximate result related to the specific choice of the experimental parameters.
It is a necessary condition for obtaining zero slopes in the middle of each plateau but not a sufficient condition for obtaining strict flatness over the whole range of each plateau.
In the case of non-overlapping constituent patterns, i.e. $\Delta z\gg 2\sigma$, the third condition for rigidity simplifies to $\Delta z=$constant. In this case we can also relax (in an arbitrary system that does not inherently satisfy the first two conditions) the other conditions: it is sufficient that $\Delta\theta$ is a constant (but not necessarily zero) and $\kappa\sigma$ is not required to be conserved.
In this case the AFT of the pair of patterns
$|{\cal F}V^{(+)}(K)|=e^{-\frac12\sigma^2(K-\kappa)^2}|\cos(\frac12 K\Delta z-\Delta\theta)|$
is a product of a Gaussian dependent of $\kappa$ and a cosine that does not vary when $\kappa$ is scanned. This AFT is an absolute cosine with Fourier space periodicity $2\pi/\Delta z$ with a Gaussian envelope whose width $\kappa$ and extends over $1/\sigma$ extends over a few periods of the cosine and its center varies with $\kappa$. When $\kappa$ is changed the peaks of the cosine stay static and $K_M$, defined as the largest peak of the AFT, changes only when the peak of the Gaussian function is closer to the next cosine peak.

This model may be implemented by analyzing the $\Delta\theta=$constant output of a mode-locked laser whose output is a train of light pulses at constant delay times and a constant phase shift (so-called carrier-envelope offset) between each pulse and the next one. By analyzing the Fourier transform of pairs of consequent pulses of this laser we can obtain exactly the same result shown here: the main peak of the spectrum of such a pair will show exactly the same behavior of $K_M$ in Fig.\,\ref{fig:universal}, where its value is constant within a range of frequencies, giving rise to a quantized spectrum of $K_M$. Obviously, as noted previously, such tests may only be done with frequencies for which available detectors have enough bandwidth to follow the oscillations.

\section{Phase gradient analysis}

In this section we provide a more intuitive insight into the effects presented in this work by looking at the variation of the phase $\varphi$ of the moire pattern in real space.
Consider two infinite sinusoidal periodic patterns with the same wavenumber and amplitude but different phases $\phi_1$ and $\phi_2$. Their sum is a similar periodic pattern with the same periodicity:
$\cos(kx+\phi_1)+\cos(kx+\phi_2)=A\cos[\varphi(z)]$, where $A=2\cos\left[\frac12(\phi_1-\phi_2)\right] $ and $\varphi(z)=kz+\frac12(\phi_1+\phi_2)$. The periodicity can be defined as the phase gradient $\partial\varphi/\partial z=k$. If the amplitudes of the two patterns are not equal, then the sum pattern phase is closer to the phase of the pattern with the larger amplitude. Furthermore, if these amplitudes vary with $z$, then the sum pattern's local phase is not linear in $z$, and one can define a local periodicity $\partial\varphi/\partial z=k+\delta k(z)$.

The positive frequency part of the sum pattern in Eq.~(1) of the main text can be written as a complex function with a pre-factor $e^{i\kappa z}$ and a sum of two translated Gaussians with two $z$-independent phase factors [Eq.~(3) of the main text]
\begin{equation}
V^{(+)}(z)=e^{i\kappa z}[G_-(z)e^{i\tilde{\phi}_1}+G_+(z)e^{i\tilde{\phi}_2}],
\end{equation}
with $G_{\pm}(z)=\exp[-(z-\bar{z}\pm \Delta z/2)^2/2\sigma^2]$, $\bar{z}\equiv \frac12(z_1+z_2)$and $\tilde{\phi}_j=2\theta_j-\kappa z_j$.
The local phase of this function in the complex plane is
\begin{equation}
\varphi(z)=\kappa z+{\rm atan}\left[\frac{G_-(z)\sin\tilde{\phi}_1+G_+(z)\sin\tilde{\phi}_2}{G_-(z)\cos\tilde{\phi}_1+G_+(z)\cos\tilde{\phi}_2}\right].
\end{equation}
We define $\Delta\phi=\tilde{\phi}_1-\tilde{\pi}_2$ and obtain
\begin{equation}
\varphi(z)= \\ \kappa z+
\tilde{\phi}_1-{\rm atan}\left[\frac{\sin\Delta\phi}{e^{-(z-\bar{z})\Delta\phi/\kappa\sigma^2}+\cos\Delta\phi}\right],
\end{equation}
where $\Delta\phi=\kappa\Delta z$ if $\theta_1=\theta_2=\phi$ in Eq.~(1) of the main text.

\begin{figure}[h]
\centerline{
\includegraphics[trim={0cm 0 0cm 0cm},clip,width=0.8\columnwidth]{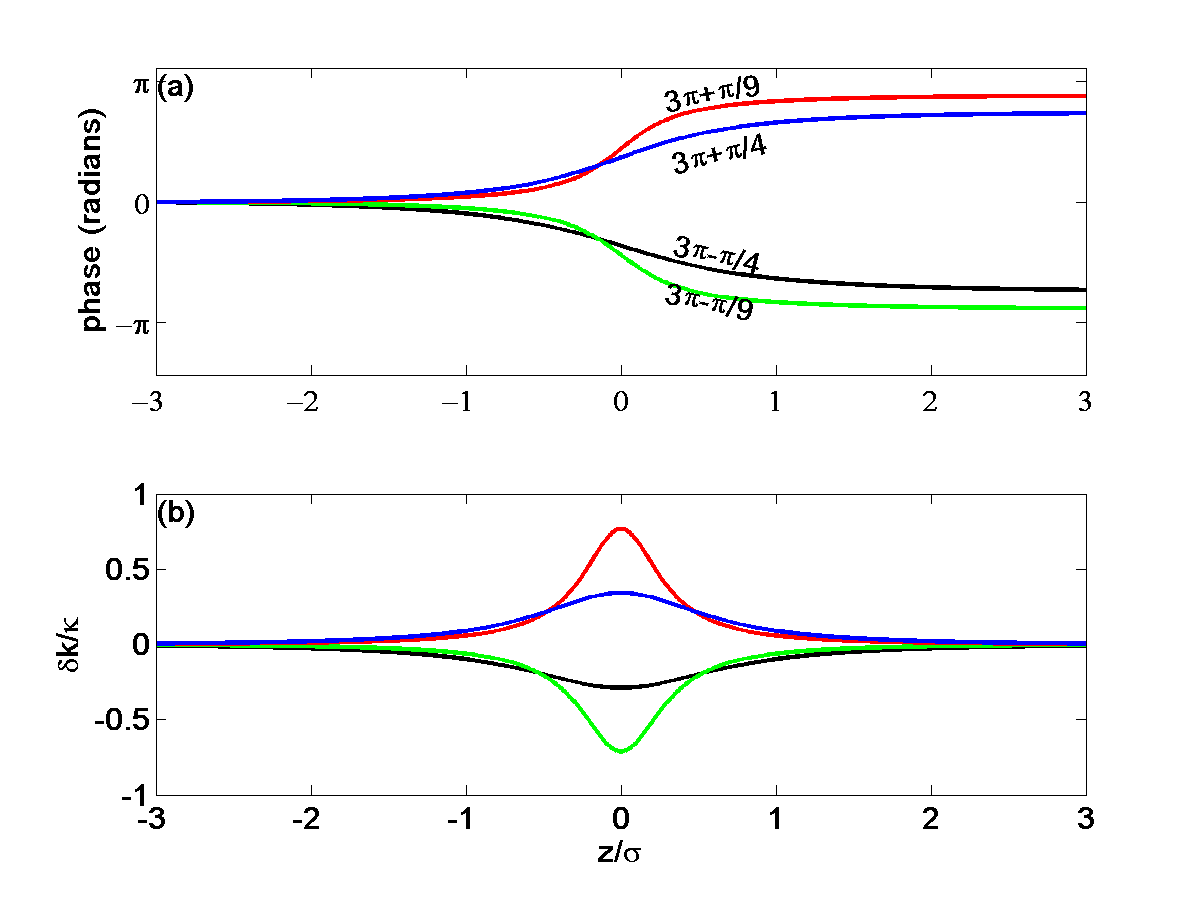}}
\caption[]{Phase variation of a moire pattern along $z$ relative to the phase $\varphi_1(z)=\kappa (z-z_1)$ of one of its constituents. The phase (a) changes from $\varphi_1$ (left) to $\varphi_2=\varphi_1-\Delta\phi+2\pi n$ (right), where $ n$ is the integer closest to $\Delta\phi/2\pi$, which is smaller than $\Delta\phi$ for the lower curves ($2\pi<\Delta\phi<3\pi$) and larger for the upper curves ($3\pi<\Delta\phi<4\pi$). The phase gradient (b) is therefore negative in the transition region between the dominance regimes of the two constituent patterns when $\Delta\phi<3\pi$  and positive when $\Delta\phi>3\pi$, such that when $\Delta\phi$ is scanned through an odd multiple of $\pi$ the deviation of the wavenumber $K_M$ from $\kappa$ jumps from a minimal negative value to a maximal positive value of the next order (integer $n$).}
\label{fig:phasegrad}
\end{figure}

The local gradient of the phase is then
\begin{equation}
\frac{\partial\varphi}{\partial z}=\kappa\left(1-\frac{\Delta\phi}{2(\kappa\sigma)^2} \frac{\sin\Delta\phi}{{\{\rm cosh}\left[\frac{(z-\bar{z}) \Delta\phi}{\kappa\sigma^2}\right] +\cos\Delta\phi}\right)
\end{equation}
The phase variation is demonstrated in Fig.~\ref{fig:phasegrad}. In the transition region between the dominance ranges of the two constituent patterns the phase gradient, which represents an additional effective contribution to the wave number, reduces when $\Delta\phi$ is larger than the closest value of $2\pi n$ and increases when $\Delta\phi$ is smaller then the closest $2\pi n$. This gives rise to a jump of $K_M$ at $\Delta\phi=\pi(2n+1)$.

\begin{figure}
\includegraphics[width=\columnwidth]{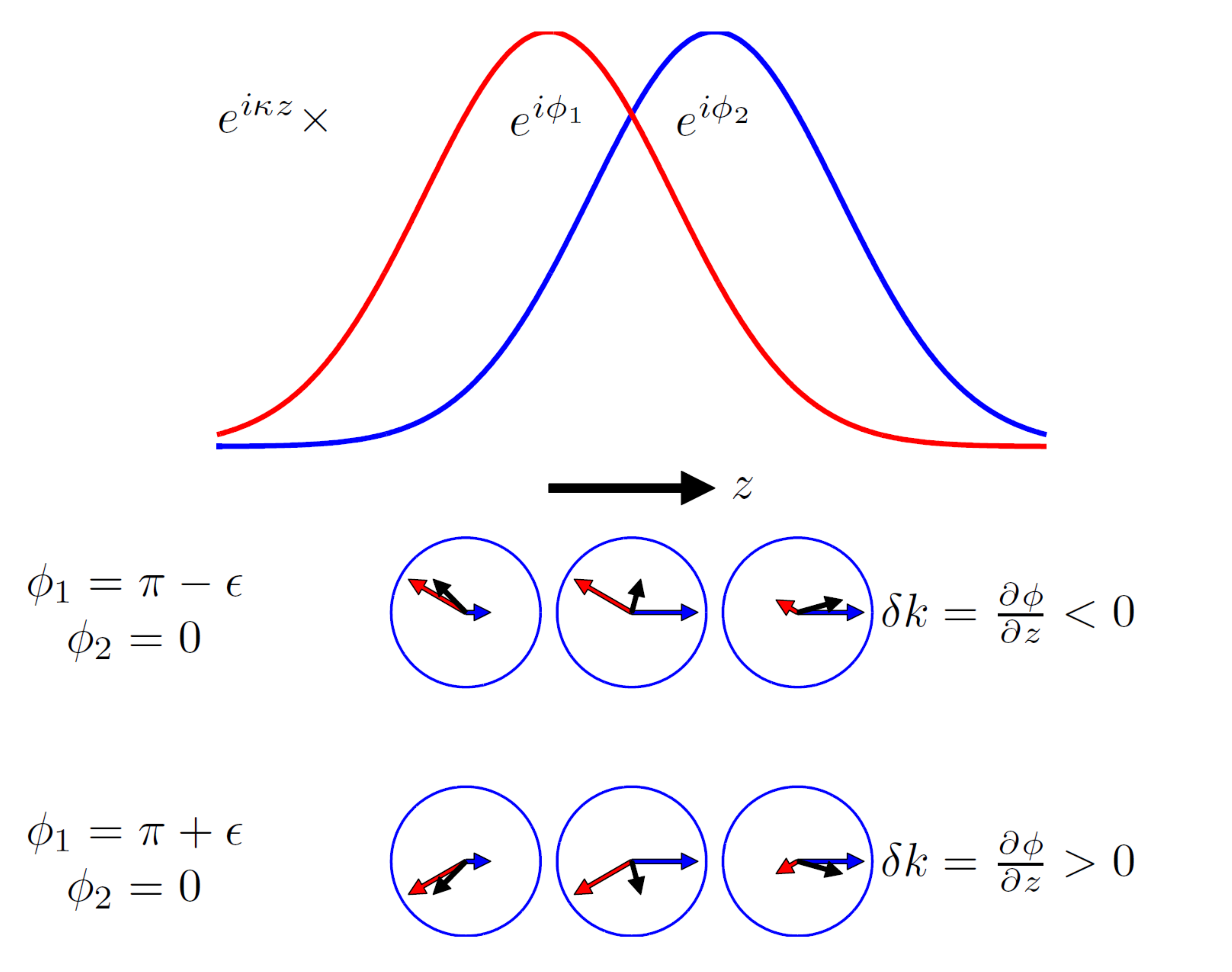}
\caption{Real-space explanation for the shift and jumps of the moire pattern periodicity $K_M$. The local phase in a sum of two Gaussian envelopes centered at $z_1$ and $z_2$ ($z_1<z_2$) with global phases $\phi_1$ and $\phi_2$, respectively. The phase of the sum pattern is closer to $\phi_1$ when $z<\frac12(z_1+z_2)$ and closer to $\phi_2$ when $z>\frac12(z_1+z_2)$. It follows that if $\Delta\phi=\phi_1-\phi_2$ is in the range $0<\Delta\phi<\pi$ (upper line of circles) the phase decreases along $z$ (from about $\phi_1$ to about $\phi_2<\phi_1$), corresponding to a negative contribution $\delta k=\partial\phi/\partial z<0$ to the wavenumber $K_M$. If $\pi<\Delta\phi<2\pi$, or equivalently $-\pi<\Delta\phi<0$ the phase increases along $z$ (lower line of circles) so that the contribution to $K_M$ is positive $\delta k>0$. When the phase difference $\Delta\phi$ is scanned through $\pi$ the periodicity wave-vector jumps from below $k_f$ to above $k_f$.}
\label{fig:phasesum}
\end{figure}

For a qualitative explanation of the effect see also Fig.~\ref{fig:phasesum}.

\section{Numerical simulation}
\label{sec:numerical}
The numerical simulation presented in Fig.\,2 of the main text, was performed by using the wavepacket evolution method \cite{Margalit2019,Japha2019}
For a BEC under the influence of time-dependent potentials. After the final gradient pulse of duration $T_3$, there are four wavepackets such that each pair corresponding to the same spin state is summed coherently to yield an interference pattern, and the probabilities of the two interference patterns are summed incoherently to yield the moire pattern. The initial state is a BEC of $10^4$ $^{87}RB$ atoms in a trap of frequencies $\omega_x=2\pi\times 38$\,Hz and $\omega_y=\omega_z=2\pi\times 113$\,Hz. Taking into account the number of atoms and the $s$-wave scattering length of $^{87}$Rb ($=100\times $ Bohr radius) we obtain an atomic density with sizes (in terms of standard deviation) $\sigma_x=3.45\,\mu$m and $\sigma_y=\sigma_z=1.23\,\mu$m.
Note that these calculated numbers are for a pure BEC model with the given number of atoms. The wavepacket sizes in the trap are not measurable directly. The final clouds' measured sizes after the formation of the interference fringes turn out to be larger than the prediction of the numerical model but almost 20\%. This discrepancy may be attributed to effects that are not accounted for in the simulation, such as an incomplete condensation in the trap or unknown magnetic field gradients during the trap release.

\begin{figure}
\includegraphics[width=\columnwidth]{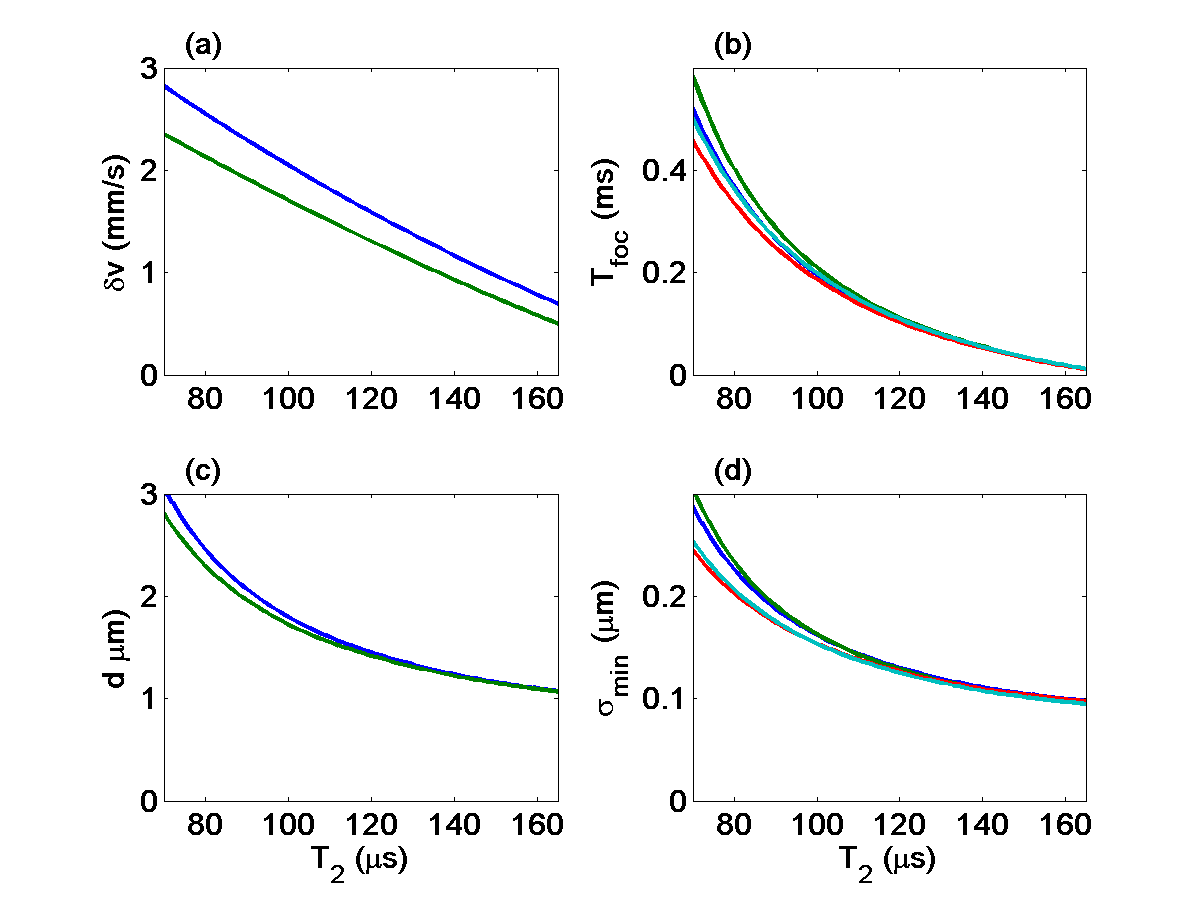}
\caption{Calculated properties of the wavepackets after the final gradient pulse of duration $T_3$ as a function of the deceleration pulse duration $T_2$. (a) momentum difference between the two wavepackets of the same spin state (the upper curve is for the state $|1\rangle$). (b) The time between the end of the final gradient pulse and the focusing time where the wavepackets' size is minimal (the different curves correspond to the 4 wavepackets). (c) Distance between the wavepackets of the same spin at the focusing point. (d) Minimal wavepacket size for the four wavepackets.}
\label{fig:wppars}
\end{figure}

The simulation used the experimental parameters. Some of the parameters are not known exactly, and we choose their values such that the numerical calculation agrees with the experimental measurements, particularly with the periodicities of the interference patterns of each of the spin states ($\kappa_1$ and $\kappa_2$). We take the initial trapping distance from the chip to be $z_0=89.5\,\mu$m, the duration of the splitting gradient pulse to be $T_1=3.75\,\mu$s and the chip current during the gradient pulses to be $I=1.122$\,A. These values are well within the experimental uncertainty, and they reproduce quite accurately the values of $\kappa$ that are measured independently. In addition, to reproduce the positions of the spatial frequency jumps in Fig.~2 of the main text, we take the bias gradient during the whole evolution (except the last 13\,ms when the bias is turned off) to be 90 G/m. This value is not completely supported by experimental evidence, but it turns out to be the right value to consider for a few possible sources of the magnetic field's inhomogeneity.
Fig.~\ref{fig:wppars} shows some properties of the wavepackets during their evolution, which are not directly measured in the experiment.

\section{ Demonstration of the conservation laws in phase space}
\label{sec:Wigner}

In this section we complement the proof of the conservation laws in previous sections by a demonstration that indicates that the conservation of the number of interference fringes and their phase do not essentially require the overlap integral to be involved.
While the proof given above is general in the sense that it shows that the quantities $\Gamma$ and $\chi$ are conserved in any unitary operation, the specific form of $\Gamma$ is in terms of parameters of a Gaussian wavepacket and it relies on the overlap integral between the two wavepackets, which may crucially depend on details of the shape of the wave functions at their tails. Here we present a complementary vision that does not rely on an overlap integral and may apply to wavepackets that are completely non-overlapping. On the other hand, it is based on the assumption that the potential acting during the evolution can be fairly well described by a quadratic form over the region occupied by the distribution.

\begin{figure}
\includegraphics [width=.45\textwidth]{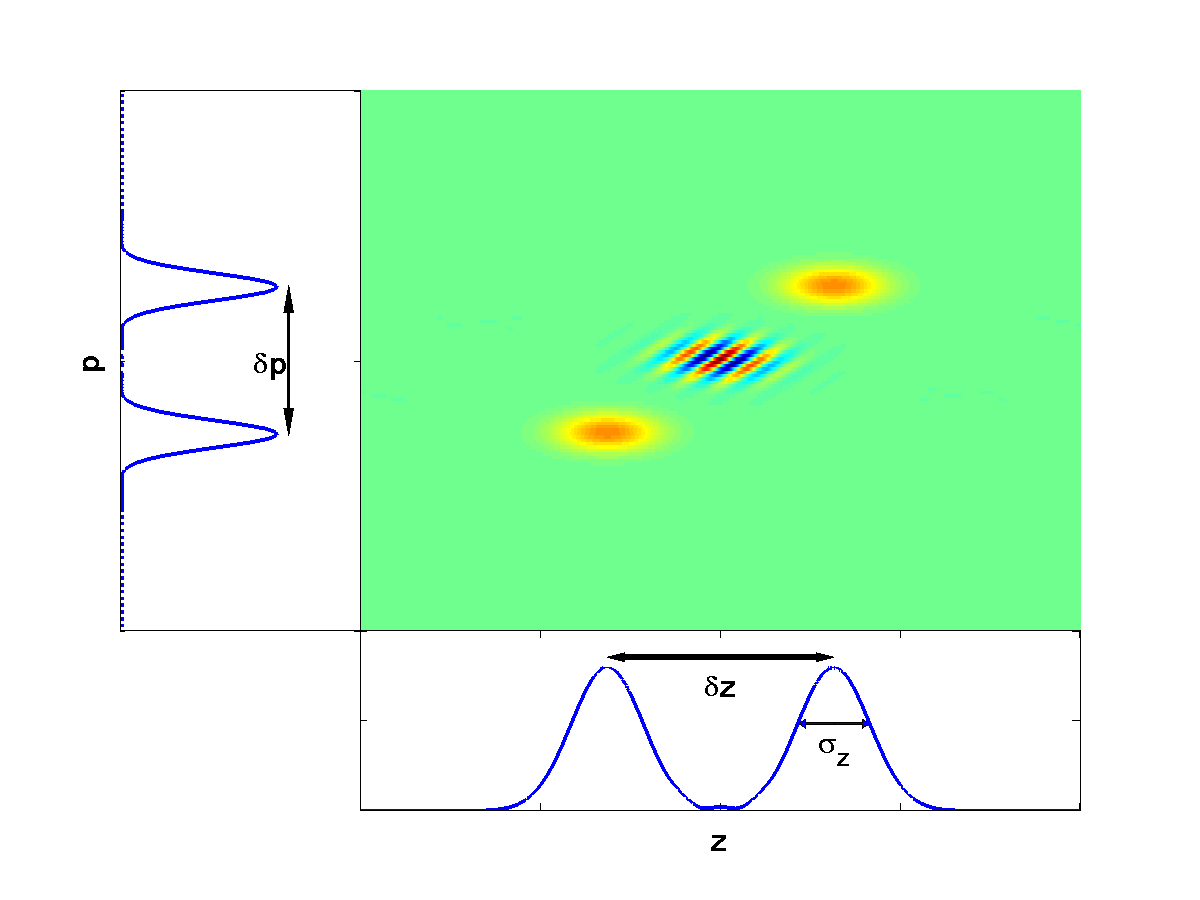}
\caption{Wigner function phase-space density of a superposition of two Gaussian wave-packets separated by $\delta p$ and $\delta z$. The shape of the pattern in phase space is conserved under phase space rotations and axis scaling operations.} \label{fig:Wigner}
\end{figure}

As we have shown in a previous work that analyzed the interferometric sequence in our Stern-Gerlach interferometer~\cite{Margalit2019}, the evolution of the pair of wavepackets can be viewed as a scaled phase space rotation (see Fig.~9 in Ref.~\cite{Margalit2019} and an additional rigorous proof here below). The form of the phase space distribution of a superposition of two wavepackets consists of two peaks at the phase space coordinates where these wavepackets are centered and a fringe pattern that appears in the Wigner distribution in between these centers, whose wave vector points perpendicular to the line that connects between the phase space center coordinates, as demonstrated in Fig.~\ref{fig:Wigner}. Under rotation of phase space coordinates this structure conserves the form of this fringe pattern and in particular it conserves the number of fringes along this pattern and the phase, namely the position of the fringes relative to the center of the pattern. When the two coordinate centers are separated by momentum, while the distributions around these centers overlap in space, the fringe pattern appears as a real interference pattern in space. On the other hand, when the two centers are separated in space, such as during the focusing of the wavepackets before they expand, then the interference fringes do not appear in the real space distribution but rather in the momentum distribution.

Let us now explicitely derive the starting point result that evolution in a quadratic potential can be represented by a phase space rotation.
The Wigner function for a pure state represented by a wave function $\psi(x,t)$ in one dimension is defined as
\begin{equation}
W(x,p)=\frac{1}{2\pi\hbar}\int_{-\infty}^{\infty} d\eta\, e^{-i\eta p/\hbar}\psi(x+\frac{\eta}{2})\psi^*(x-\frac{\eta}{2}).
\end{equation}
If $\psi(x,t)$ satisfies the Schr\"odinger equation $i\partial\psi/\partial t=-(i/\hbar)\hat{H}\psi$, where $\hbar{H}$ is the Hamiltonian, then the evolution of the Wigner function is determined by
\begin{equation}
 \frac{\partial W(x,p)}{\partial t}=-\frac{i}{\hbar}\int_{\eta}\left\{ [\hat{H},\psi(x+\frac{\eta}{2})\psi^*(x-\frac{\eta}{2})]\right\}_p,
\end{equation}
where for brevity we define
\[  \int_{\eta} \{\dots\}_p\equiv \frac{1}{2\pi\hbar}\int_{-\infty}^{\infty}d\eta\,e^{-i\eta p/\hbar}\dots. \]
and the Hamiltonian in the commutation relation operates only on the wae fnction closest to it.

Let us now assume a Hamiltonian with a quadratic potential $\hat{H}=\hat{p}^2/2m+\frac12 m\omega^2\hat{x}^2$.
The kinetic part of the Hamiltonian $\hat{p}^2/2m$, where $\hat{p}=-i\hbar\partial/\partial x$ acts as
\begin{eqnarray}
\left. \frac{\partial W}{\partial t}\right|_{\rm kin}& = & \frac{i\hbar}{2m}\int_{\eta}\left\{ \psi''(x+\frac{\eta}{2})\psi^*(x-\frac{\eta}{2}) \right. \nonumber \\
&& \left. -\psi(x+\frac{\eta}{2})[\psi''(x-\frac{\eta}{2})]^*\right\}_p \,.
\end{eqnarray}
When integrating by parts the terms involving  $\psi'(x+\frac{\eta}{2})\psi'(x-\frac{\eta}{2})^*$ cancel and we are left with the term
\begin{eqnarray}
&=& -\frac{p}{m}\int_{\eta}\{ \psi'(x+\frac{\eta}{2})\psi^*(x-\frac{\eta}{2})+\psi(x+\frac{\eta}{2})\psi'(x-\frac{\eta}{2})^*]\}_p \nonumber \\
& =& -\frac{p}{m}\frac{\partial W}{\partial x}
\end{eqnarray}
Similarly a harmonic potential term $\frac12 m\omega^2\hat{x}^2$ in the Hamiltonian acts as
\begin{eqnarray}
\left.\frac{\partial W}{\partial t}\right|_{\rm pot} &=& - \frac{i}{2}m\omega^2\int_{\eta}\{[(x+\frac{\eta}{2})^2-(x-\frac{\eta}{2})^2] \nonumber \\
&&\times \psi(x+\frac{\eta}{2})\psi^*(x-\frac{\eta}{2})\}_p.
\end{eqnarray}
As $(x+\frac{\eta}{2})^2-(x-\frac{\eta}{2})^2=2x\eta=2i\hbar x(\partial/\partial p)$
we find that this expression is  equal to $m\omega^2x\partial W/\partial p$. It follows that the equation for the Wigner function is
\begin{equation}
\frac{\partial W(x,p)}{\partial t}=-\frac{p}{m}\frac{\partial W}{\partial x}+m\omega^2 x\frac{\partial W}{\partial p}.
\end{equation}
For an infinitesimal time interval $\delta t$ the solution is
\begin{equation}
 W(x,p,t+\delta t)=W(x-\frac{p}{m}\delta t,p+m\omega^2 x\delta t,t),
\end{equation}
while for an arbitrary time interval $\tau$ this becomes
\begin{eqnarray}
&& W(x,p,t+\tau)= \nonumber \\
&& W(x\cos\omega\tau-\frac{p}{m\omega}\sin\omega\tau,p\cos\omega\tau +m\omega x\sin\omega\tau,t),
\end{eqnarray}
We have then proved that evolution in a quadratic Hamiltonian is equivalent to phase space rotation of the Wigner function of an arbitrary pure state. This proof can be easily extended to impure states that are given by a density matrix which is a weighted sum over density matrices of pure states.

%
%
\section*{Extended Data}

\vspace{1 cm}

\clearpage

\begin{thebibliography}{99}

\bibitem{review}  V. Saveljev,  J. Kim, J. Y. Son, Y. Kim, G. Heo, Static moir\'e patterns in moving grids. {\it Sci. Rep.} {\bf 10}, 14414 (2020), and references therein.

\bibitem{Zhang2017} C. Zhang et al. Interlayer couplings, moir\'e patterns, and 2D electronic superlattices in MoS2/WSe2 hetero-bilayers. {\it Sci. Adv.} {\bf 3}, e1601459 (2017).

\bibitem{Chen2020} X. Chen et al. Moir\'e engineering of electronic phenomena in correlated oxides. {\it Nat. Phys.} {\bf 16}, 631-635 (2020).

\bibitem{Tran2019} K. Tran et al. Evidence for moir\'e excitons in van der Waals heterostructures. {\it Nature} {\bf 567}, 71-75 (2019).

\bibitem{Kennes2021} D. M. Kennes et al. Moir\'e heterostructures as a condensed-matter quantum simulator. {\it Nat. Phys.} {\bf 17}, 155–163 (2021).

\bibitem{keil} M. Keil et al. Fifteen years of cold matter on the atom chip: promise, realizations, and prospects. {\it J. Mod. Opt.} {\bf 63:18}, 1840-1885 (2016).











\bibitem{Ashmead2012} J. Ashmead, Morlet wavelets in quantum mechanics. {\it Quanta} {\bf 1}, 58-70 (2012).




\bibitem{Jin2019} C. Jin et al. Observation of moir\'e excitons in WSe2/WS2 heterostructure superlattices. {\it Nature} {\bf 567}, 76-80 (2019).

\bibitem{Alexeev2019} E. M. Alexeev et al. Resonantly hybridized excitons in moir\'e superlattices in van der Waals heterostructures. {\it Nature} {\bf 567}, 81-86 (2019).

\bibitem{Brotons2020} M. Brotons-Gisbert et al. Spin–layer locking of interlayer excitons trapped in moir\'e potentials. {\it Nat. Mater.} {\bf 19}, 630-636 (2020).

\bibitem{Hunt2013} B. Hunt et al. Massive Dirac Fermions and Hofstadter Butterfly in a van der Waals Heterostructure. {\it Science} {\bf 340}, 1427-1430 (2013).

\bibitem{Dean2013} C. R. Dean et al. Hofstadter’s butterfly and the fractal quantum Hall effect in moir\'e superlattices. {\it Nature} {\bf 497}, 598-602 (2013).

\bibitem{Ponomarenko2013} L. A. Ponomarenko et al. Cloning of Dirac fermions in graphene superlattices. {\it Nature} {\bf 497}, 594-597 (2013).

\bibitem{Shi2014} Z. Shi et al. Gate-dependent pseudospin mixing in graphene/boron nitride moir\'e superlattices. {\it Nat. Phys.} {\bf 10}, 743-747 (2014).

\bibitem{Gorbachev2014} R. V. Gorbachev et al. Detecting topological currents in graphene superlattices. {\it Science} {\bf 346}, 448-451 (2014).

\bibitem{Song2015} J. C. W. Song, P. Samutpraphoot, L. S. Levitov, Topological Bloch bands in graphene superlattices. {\it PNAS USA} {\bf 112}, 10879-10883 (2015).

\bibitem{Spanton2018} E. M. Spanton et al. Observation of fractional Chern insulators in a van der Waals heterostructure. {\it Science} {\bf 360}, 62-66 (2018).

\bibitem{Deilmann2020}  T. Deilmann, M.  Rohlfing, U.  Wurstbauer,  Light–matter interaction in van der Waals hetero-structures. {\it J. Phys.: Condens. Matter} {\bf 32}, 333002 (2020).

\bibitem{Bistritzer2011} R. Bistritzer, A. H. MacDonald, Moir\'e bands in twisted double-layer graphene. {\it PNAS USA} {\bf 108}, 12233-12237 (2011).

\bibitem{Cao2018}  Y. Cao et al. Unconventional superconductivity in magic-angle graphene superlattices. {\it Nature} {\bf 556}, 43-50 (2018).

\bibitem{Cao2018b} Y. Cao et al. Correlated insulator behaviour at half-filling in magic-angle graphene superlattices. {\it Nature} {\bf 556}, 80-84 (2018).

\bibitem{Stepanov2020} P. Stepanov et al. Untying the insulating and superconducting orders in magic-angle graphene. {\it Nature} {\bf 583}, 375-378 (2020).

\bibitem{Nuckolls2020} K. P. Nuckolls et al. Strongly correlated Chern insulators in magic-angle twisted bilayer graphene. {\it Nature} {\bf 588}, 610–615 (2020).

\bibitem{Zondiner2020} U. Zondiner et al. Cascade of phase transitions and Dirac revivals in magic-angle graphene. {\it Nature} {\bf 582}, 203-208 (2020).

\bibitem{Andrei2020} E. Y. Andrei, A. H. MacDonald, Graphene bilayers with a twist. {\it Nat. Mater.} {\bf 19}, 1265–1275 (2020).

\bibitem{Margalit2019} Margalit, Y. et al. Analysis of a high-stability Stern-Gerlach spatial fringe interferometer. {\it New J. Phys.} {\bf 21}, 073040 (2019).

\bibitem{Margalit2018} Margalit, Y. et al. Realization of a complete Stern-Gerlach interferometer: Towards a test of quantum gravity, {\it Science Advances}, in print (2021). https://arxiv.org/abs/2011.10928 (2020).

\bibitem{Japha2019} Japha, Y. Generalized wave-packet model for studying coherence of matter-wave interferometers. https://arxiv.org/abs/1902.07759 (2020).


\end{thebibliography}
\end{document}